\documentclass[sigplan,10pt]{acmart}
\renewcommand\footnotetextcopyrightpermission[1]{}

\AtBeginDocument{%
  }

\acmSubmissionID{\#1329}
\copyrightyear{2026}
\acmYear{2026}
\acmConference[ATC]{ACM SIGOPS Annual Technical Conference}{November 15--18}{Hong Kong, China}
\usepackage[]{hyperref}

\usepackage{tikz}
\usepackage{amsmath}
\usepackage{xurl}
\usepackage{enumitem}
\usepackage{multirow}
\usepackage{booktabs}
\usepackage{pifont}
\usepackage{subfig}
\usepackage{float}
\usepackage{diagbox}
\usepackage{threeparttable}
\usepackage{balance}
\usepackage[]{hyperref}
\usepackage{colortbl}
\usepackage{xcolor}
\usepackage{todonotes}
\usepackage{amsmath}
\usepackage{diagbox}
\usepackage{comment}
\usepackage[linesnumbered,ruled,vlined]{algorithm2e}
\usepackage[shortcuts]{extdash}

\newcommand{\sys}{\textit{EnerInfer}\xspace}

\definecolorseries{test}{rgb}{grad}[rgb]{.95,.55,.55}{11,11,17}
\resetcolorseries[10]{test}
\newcommand{\addtodoeditor}[1]{%
    \colorlet{#1}{test!!+!50}
    \expandafter\newcommand\csname#1\endcsname [1]{%
        \todo[color=#1,size=\tiny]{\sffamily\textbf{\uppercase{#1}:}
    ##1}\xspace%
    }
    \expandafter\newcommand\csname#1i\endcsname [1]{%
        \todo[inline, color=#1, size=\tiny]{\sffamily\textbf{\uppercase{#1}:} ##1}\xspace%
    }
}
\addtodoeditor{dr}
\addtodoeditor{bs}
\addtodoeditor{bz}

\newcommand{\revise}[1]{
{
#1}
}

\usepackage{caption}
\begin{document}

\title{EnerInfer: Energy-Aware On-Device LLM Inference}

\author{
Bohua Zou$^{1,2}$, 
Nian Liu$^{4}$,
% Yu Wang$^{1}$,
Binqi Sun$^{1}$,
Matteo Mascherin$^{2}$,
Debayan Roy$^{2}$,
% Diogo Behrens$^{2}$,
Yutao Liu$^{2}$, 
Yu Peng$^{3}$, \\
Ning Jia$^{3}$, 
Haibo Chen$^{3,4}$}

\affiliation{%
\vspace{1ex}
  \institution{$^{1}$Technical University of Munich\hspace{0.6cm}$^{2}$Huawei Hilbert Research Center (Dresden)\hspace{0.6cm}$^{3}$Huawei Central Software Institute\hspace{0.6cm}$^{4}$Shanghai Jiao Tong University}
  \city{}
  \state{}
  \country{}
  % \vspace{0.5ex}
}

\renewcommand\shortauthors{}
\renewcommand\shorttitle{}
\begin{abstract}
On-device LLM inference is increasingly attractive for privacy-preserving, reliable, and cost-effective deployment, yet its energy and thermal costs remain a critical bottleneck. Existing systems primarily optimize for decoding speed, implicitly assuming that faster execution is always preferable. We show instead that on-device LLM inference often has exploitable configuration slack: modestly lowering NPU and memory frequencies preserves quality of experience (QoE) while substantially improving energy efficiency and reducing heat. 

Realizing this opportunity in production is challenging. The most energy-efficient NPU/DDR setting varies with the model, inference engine, platform, and runtime conditions, with no stable ranking across configurations. Commercial devices further lack component-level power sensing, and shell temperature evolves with request arrivals, response lengths, and thermal history.
To address these challenges, we propose {\sys}, the first on-device LLM inference framework that jointly manages energy efficiency, throughput, and thermal comfort for LLM workloads. {\sys} replaces per-model profiling and sensor-heavy control with disaggregated, model-structure-aware prediction and ranking-driven online feedback. It predicts throughput and power for unseen LLMs across NPU/DDR frequency settings, selects QoE-satisfying efficient configurations under runtime interference, and uses lightweight limited-horizon thermal prediction to dynamically switch between energy-optimized and thermally constrained inference.
Evaluations on real-world LLMs show that {\sys} improves energy efficiency by up to 65\%, 12\%, and 24\% on phones, a laptop, and a development board, respectively, without QoE violation.

\end{abstract}

% \begin{IEEEkeywords}
% Article submission, IEEE, IEEEtran, journal, \LaTeX, paper, template, typesetting.
% \end{IEEEkeywords}

% \keywords{Power Efficiency, LLM}
\setcopyright{none}
\maketitle
\pagestyle{plain}

\begingroup
\renewcommand\thefootnote{}
\footnotetext{
Corresponding author: Nian Liu,
\href{mailto:nianliu@sjtu.edu.cn}{nianliu@sjtu.edu.cn};
}
\addtocounter{footnote}{-1}
\endgroup

\section{Introduction}

Large language models (LLMs) have garnered widespread attention since their inception,
powering a variety of applications, from conversational agents to tasks such as text polishing and translation.
On-device LLM inference has become a focal point of academic research~\cite{pi, HeteroLLM, llmnpu, MELTing, personalllm} and is increasingly getting adopted by industry~\cite{apple-on-device, pangupi}, as it improves privacy by keeping data local, offers low and stable latency, while reducing vendor costs.
Even as on‑device LLMs achieve substantial gains in prefill and decoding speed~\cite{llmflash, pi, llmnpu}, their high and insufficiently studied power consumption continues to erode battery life and heighten battery anxiety, undermining their practicality in everyday mobile scenarios.
%removed jiang2024d

\begin{figure}[t]
    \centering
    \includegraphics[width=\linewidth]{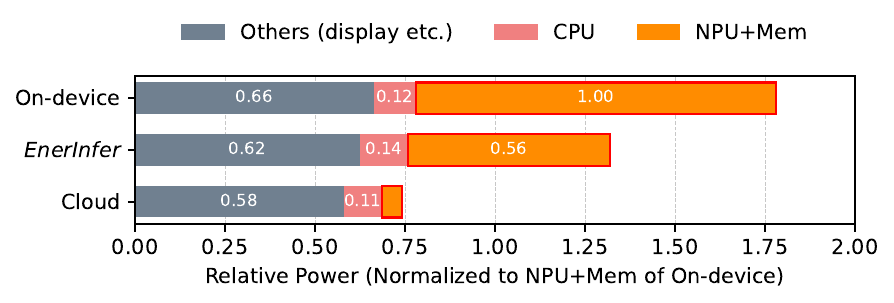}
    \Description{A bar plot of the breakdown energy consumption of three different types of energy consumption.}
   %  \caption{Comparison of on-device vs. cloud-offloaded inference for text polishing on a phone.
   % On-device inference incurs high power.
   % {\sys} cuts power by 26\%, improves energy efficiency by 38\% (tokens/joule, w.r.t. \emph{NPU+Mem}, as \emph{Others+CPU} remain stable and unrelated to inference), while preserving QoE (10 t/s) and incurring no accuracy loss.}
   \caption{Component-wise power consumption of LLM-based text polishing on a phone under the default settings and our method for on-device inference, as well as a cloud-offloaded inference.}
   \label{fig:moti}
   \vspace{-4mm}
    % \bzi{This is the old data, should we change that?}
\end{figure}

Figure~\ref{fig:moti} illustrates the normalized average power in a text-polishing scenario using an LLM~\cite{pangupi} on a phone.
As shown, the default on-device inference consumes more than 138\% of the power required by cloud offloading.
Most of this additional power is consumed by the neural processing unit (NPU) and memory (Mem).
% ---the primary components engaged during an inference.
% which operate at the maximum frequencies to maximize the throughput (i.e., speed or tokens-per-second, t/s).
% due to the high compute and memory-bandwidth demands during inference, 
\revise{Because the \emph{decode phase} is heavily memory‑bandwidth‑bound and the NPU offers limited dynamic voltage and frequency scaling (DVFS) control,}
the default NPU and Mem governors run both frequencies at their maximum to maximize the decoding throughput, measured in \emph{tokens per second}.
However, an alternative frequency configuration can significantly reduce NPU and Mem power consumption by 44\% while still maintaining a throughput above 10 tokens/s.
As a result, the energy efficiency---measured in \emph{tokens per joule} of NPU and Mem energy---improves by 38\%.
%thereby improving energy efficiency---measured in \emph{tokens per joule}---by 38\%.
% Moreover, it causes a noticeable temperature rise on the phone’s back shell ($4.2^{\circ}\mathrm{C}$ in a minute), leading to user discomfort and thermal throttling after a short execution.
% Given the impressive speeds achieved~\cite{pi2, llmnpu, jiang2024d} and the rapidly growing demand for on-device inference, it is now crucial to address the overlooked energy issue.
%
%
This highlights an opportunity that \emph{many scenarios do not require maximum throughput}, and \emph{reducing NPU and Mem frequencies can significantly improve energy efficiency, while still maintaining a satisfactory quality of experience (QoE) and incurring no accuracy loss.}
For instance, in a voice assistant scenario, speaking is typically $<$3 words/s~\cite{speakingspeed}, and reading $<$5.3 words/s~\cite{readspeed-jornal, readingspeed}, making higher throughput unnecessary. 
A recent study also found that users read/listen at 4.8/3.3 tokens/s on average across age and language groups~\cite{Andes-QoE}.

 % esweek

In this work, we first gain a comprehensive understanding of the energy efficiency of on-device LLM inference through \emph{extensive analysis} of throughput and power consumption under different hardware configurations (i.e., NPU and Mem frequencies) for a wide range of \emph{synthetic} LLMs---self-engineered by varying hyperparameters---as well as production\-/grade LLMs (\S\ref{sec:observation}).
Our experiments span \emph{diverse platforms}, including a smartphone, a laptop, and a development board~\cite{orangepi_5pro}. 
Based on \emph{our experimental analysis}, we list the following challenges towards optimizing the energy efficiency while meeting the throughput target.
\begin{itemize}[leftmargin=*]
% \begin{itemize}
\item $\mathcal{C}_1$.~Both energy efficiency and throughput under a specific hardware configuration vary significantly across models and inference engines, and these variations are also platform-dependent.
% Given the rapid evolution of LLMs---evidenced by Hugging Face recently surpassing 1 million models~\cite{hfmodels}---profiling each model offline individually is impractical in deployment.
With LLMs evolving rapidly---Hugging Face now hosts over 1 million models~\cite{hfmodels}---profiling each model offline is impractical.
Also, commercial systems typically \emph{lack hardware support for measuring component-level (e.g., NPU) power consumption}; 
% hence, it becomes inherently difficult to assess the efficiency across configurations.
hence, obtaining the energy-performance profile of a new model in a production environment is not feasible. %\bz{@Debayan, can you confirm if this is sound?}
\item $\mathcal{C}_2$.~There is \emph{no consistent partial ordering} (i.e., ranking) among configurations in terms of energy efficiency across models,
rendering simple feedback mechanisms ineffective, e.g., a minimal configuration that satisfies the speed requirement may still result in a poor efficiency.
\end{itemize}

Besides energy concerns, smartphones also face thermal constraints~\cite{thermallimit} to preserve user comfort; for example, the back-shell temperature should remain below $42^{\circ}\mathrm{C}$. 
\revise{Back-to-back} inference sessions can cause the temperature to rise continuously.
The temperature evolution depends on the chosen hardware configuration, the arrival pattern of inference requests, the response lengths, and other environmental factors. 
%A potentially consecutive inference session may lead to a sustained temperature increase.
This leads to a third challenge as follows:
\begin{itemize}[leftmargin=*]
% \begin{itemize}
    \item $\mathcal{C}_3$.  Predicting the arrival pattern of inference requests and response lengths is challenging. In particular, the existing response-length prediction techniques---such as using a proxy model~\cite{aiops2024qiu}---incur unacceptable overhead, primarily because of the \emph{auto‑regressive} nature of LLM decoding.
\end{itemize}

% Previous studies have explored energy-efficient serving of deep learning (DL) models~\cite{chow2023cofris, npu-dvfs, atcGPUmiad} and LLMs~\cite{DynamoLLM, kakolyris2024slo}, primarily in the cloud.
\revise{
Previous studies have explored generic SoC energy management~\cite{crave} as well as energy-efficient serving of deep learning (DL) models~\cite{chow2023cofris, npu-dvfs, atcGPUmiad} and LLMs~\cite{DynamoLLM, kakolyris2024slo, zhang2026rethinking}.
}
However, these existing techniques cannot address the aforementioned challenges due to the following shortcomings:
% (i)~Unlike the assumptions made in~\cite{npu-dvfs, atcGPUmiad}, obtaining energy–performance profiles of primitive operators (e.g., matrix multiplications)---both analytically and experimentally---is challenging when fine‑grained observability into NPU kernel execution is unavailable.
\revise{
(i)~Coarse‑grained resource utilization measurements on mobile SoCs~\cite{crave} cannot capture the energy–throughput characteristics of LLM inference.
(ii)~Operator-level profiling-based methods~\cite{npu-dvfs, atcGPUmiad} are impractical without fine-grained observability into NPU kernel execution.
}
% (ii)~\cite{chow2023cofris, DynamoLLM, kakolyris2024slo} rely on offline profiling of inference workloads, which may be infeasible on production mobile phones or laptops---unlike on server-grade GPUs; hence, they cannot handle new models. 
\revise{
%(iii)~Methods such as\cite{chow2023cofris, DynamoLLM, kakolyris2024slo} and even mobile‑focused approaches like~\cite{zhang2026rethinking} rely on offline profiling of fixed inference workloads, which is infeasible on production mobile devices and limits their ability to adapt to unseen models.
(iii)~Even mobile‑oriented work such as \cite{zhang2026rethinking} relies on offline profiling of fixed inference workloads, which is infeasible on production devices and cannot generalize to unseen models; the same limitation applies to \cite{chow2023cofris, DynamoLLM, kakolyris2024slo}.
}
% (iii)~They target cloud serving and explore control knobs such as batch size, graphics processing unit (GPU) frequency, and model partitioning and multiplexing; however, they do not consider a coordinated control of NPU and Mem frequencies which is necessary in unified-memory architectures~\cite{codrescu2014hexagon} commonly found in edge devices.
\revise{
(iv)~Cloud-serving techniques optimize control knobs such as batch size, GPU frequency, and model partitioning and multiplexing, but do not address coordinated NPU and memory frequency scaling, which is critical for mobile devices with unified-memory architectures~\cite{codrescu2014hexagon}.
}
\revise{
(v)~Energy efficiency and thermal limits are rarely addressed jointly in LLM inference serving.
}
%\revise{
%(v)~Existing methods for mobile devices have limited generalizability and fail to consider thermal constraints~\cite{crave,zhang2026rethinking}.
%}
% \bz{We need to add existing mobile dvfs work here.}

To mitigate the limitation of existing techniques,
%To address these aforementioned challenges, 
we propose {\sys}, an energy\-/aware on\-/device inference framework (\S\ref{sec:design}) with the following contributions:

\begin{itemize}[leftmargin=*]
    \item We present the first framework that jointly considers energy consumption, thermal impact, and throughput of \emph{on-device} LLM inference. It mitigates the overlooked energy and thermal bottlenecks of mobile NPUs and memory subsystems that %existing 
    cloud-oriented approaches do not handle.

    \item Addressing $\mathcal{C}_1$, we develop an \textit{offline} prediction pipeline that learns the strong coupling between model structural characteristics and hardware frequency configurations (NPU and DDR). Using a disaggregated machine-learning design, {\sys} accurately predicts the throughput and power of \emph{unseen} LLMs across configurations, despite lacking component-level power sensors and having only limited profiling data.

    \item Addressing $\mathcal{C}_2$, we design an \textit{online}, ranking-driven feedback control algorithm that dynamically tunes NPU and DDR frequencies to maximize energy efficiency under QoE constraints. Hence, the controller is robust against interference from co-running workloads. It further incorporates a lightweight thermal-aware controller---drawing inspiration from model predictive control (MPC)---together with a limited-horizon temperature predictor, addressing $\mathcal{C}_3$, to preserve user thermal comfort for longer under rising shell temperatures. In essence, we employ a bi-modal control strategy to dynamically switch between energy-aware and thermal-aware LLM inference. 

    \item We discuss a product-ready implementation of {\sys} (\S\ref{sec:impl}) and evaluate it (\S\ref{sec:evaluation}) on commercial smartphones and laptops spanning mid-tier to high-end devices, using real-world LLM applications (e.g., text polishing and conversational agents). Across platforms, {\sys} improves energy efficiency by 9~--~65\% and reduces overall device energy consumption by up to 11\% compared to the default OS governors, with no QoE violation. 
    % \bs{Please check and update the numbers}\dr{if it is possible to refer to Sec. 5 and 6}
\end{itemize}

The rest of the paper is organized as follows. Section~\ref{sec:background} provides the necessary background. Section~\ref{sec:observation} presents the insights from observations in terms of throughput, energy, and thermal behavior. Section~\ref{sec:design} outlines our proposed framework {\sys}, and the design of each part. Section~\ref{sec:impl} describes the implementation of {\sys}. Section~\ref{sec:evaluation} evaluates {\sys} in terms of prediction accuracy, energy efficiency improvements, the effectiveness of thermal management, and real-world energy consumption optimizations. Section~\ref{sec:related} reviews relevant related works. Section~\ref{sec:conclusion} provides concluding remarks and outlines future directions.

% \section{Burning On-device LLM Inference}
\section{Background}\label{sec:background}

% \bs{I feel like this section is mixing up (1) backgrounds including on-device LLM inference and DDR DVFS, DDR+NPU (in the TODO list) etc., and (2) our empirical observations in Section 2.2. What about we separate them into two sections: first, a background section summarizing all relevant background knowledge in different subsections, i.e., (1) LLM inference (e.g., concepts like prefill and autoregressive decoding), (2) on-device hardware-related concepts (e.g., NPUs and XPUs), (3) energy and thermal management techniques like DVFS. Then, we can have a new section introducing all the empirical observations and insights that guide our design. This structure can hopefully help highlight our contributions in identifying the observations and insights.}

\subsection{Emerging on-device inference}
\label{sec:LLM}

Large Language Models (LLMs) are transformer-based machine learning (ML) models~\cite{vaswani2023attentionneed}, typically employing a decoder\-/only structure for generative tasks.
The LLM inference process follows an auto-regressive pattern: input text is first tokenized, then processed through a prefill phase, and finally decoded token by token to produce output.
% LLMs have garnered widespread attention due to their remarkable general intelligence capabilities and have demonstrated proficiency across a wide range of tasks.\dr{this last sentence is not required, it is in Sec. 1 already.}
% including natural language conversation, reasoning, translation, and task automation.
% repete with following

\smallskip
\textbf{On-device inference.}
Smartphones, as ubiquitous personal devices, are becoming one of the most suitable interfaces for users to interact with LLM-based agents.
The industry has already introduced advanced AI assistants capable of handling a wide range of tasks on mobiles~\cite{Gemini}, from conversations, text polishing, message abstraction to UI automation~\cite{personalllm}.
%remove appleintelligence
While most of these systems currently rely on cloud services to offload heavy inference tasks, the industry is increasingly adopting on-device inference~\cite{pangupi, apple-on-device} in specific scenarios due to its enhanced user privacy through local data processing, reduced reliance on cloud services, and improved accessibility and reliability.
The adoption of on-device inference is further boosted by advances in model compression, training strategies, and specialized hardware accelerators.
Specifically, techniques such as knowledge distillation enable lightweight models, such as LLaMA 3.2-1B~\cite{grattafiori2024llama3herdmodels}, to achieve accuracy comparable to larger counterparts while significantly reducing computational overhead. %~\cite{llama32}
Additionally, model quantization enhances inference efficiency by reducing memory and processing requirements with minimal accuracy degradation.
Furthermore, PowerInfer2~\cite{pi2} and llm.npu~\cite{llmnpu} leverage NPU on phones to further accelerate inference, achieving impressive speeds (e.g., 11.68 tokens/s on decoding 47B model~\cite{pi2}).
% \dr{this paragraph is more about motivating on-device inference and less on information needed to describe and solve the problem. if we need space, we can shorten this}

\smallskip
\textbf{Execution backend.}
While many open-source LLM inference engines leverage CPUs and GPUs on phones, NPUs offer significantly better energy efficiency than GPUs and are more capable of handling highly parallel computations than CPUs~\cite{llmnpu}.
Additionally, using GPUs for inference may interfere with rendering tasks, potentially degrading the user experience. 
Hence, most commercially deployed on-device inference systems rely on NPUs~\cite{apple-on-device, oppo-npu}.
Therefore, this work primarily focuses on the \emph{energy consumption of the NPU and memory} during inference.

\smallskip
\textbf{Quality of experience (QoE).}
Similar to cloud-based inference, the QoE for on-device inference is influenced by several factors. Time-to-first-token (TTFT) defines the initial response latency before the first token appears and is largely determined by the prefill phase.
Tokens-per-second (TPS) measures the rate at which tokens are generated, reflecting the system's decoding throughput. 
Further, mobile devices and laptops must ensure that the shell temperature does not exceed a certain threshold to provide a good thermal experience~\cite{thermallimit}.
Although minimizing TTFT and maximizing TPS generally enhance QoE, most users cannot effectively consume tokens at high speeds~\cite{Andes-QoE},
leaving substantial room for improving energy efficiency and thermal experience.
% \dr{Maybe, we can put one sentence about energy efficiency and why it is important before this sentence.}
% A study found that, on average, users read at a rate of 4.8 tokens/s and listen at a rate of 3.3 tokens/s~\cite{Andes-QoE}.
% across different age and language groups}.
% Therefore, there is still significant room to explore hardware configurations that satisfy QoE while also improving energy efficiency.

\subsection{Efficient inference via DVFS}
\label{sec:dvfs}
Although operating all hardware components at peak frequency can improve performance, it compromises energy efficiency because higher frequencies necessitate increased voltage~\cite{npu-dvfs}, and power consumption scales quadratically~\cite{gonzalez1997supply}.
%While running all hardware components at peak frequency may improve throughput, it reduces energy efficiency, as higher frequencies require increased voltage~\cite{npu-dvfs}, and power scales quadratically~\cite{gonzalez1997supply}.
% On-device inference is computationally intensive and therefore power-hungry.
Hence, running inference under peak hardware frequency quickly drains the battery.
A recent work~\cite{MELTing} reports that a phone can process about 500 prompts under such a setting before shutting down.
To balance performance and energy consumption, modern SoCs adopt DVFS~\cite{dvfs}, which reduces the frequency to save energy when peak performance is not required.
Besides CPUs, DVFS is supported for GPUs~\cite{gpu-dvfs}, NPUs~\cite{npu-dvfs}, and memory~\cite{lpddr5},
enabling a broader system-level optimization.

% To effectively use DVFS, several governors are defined within the OS.
Prior works on CPU DVFS governors
leverage utilization~\cite{cpu-dvfs},
workload characteristics~\cite{workload-dvfs} and task deadlines~\cite{ddl-dvfs}.
%workload-specific heuristics~\cite{workload-dvfs,learning-dvfs},
%or gradually increases frequency to meet deadlines at runtime~\cite{ddl-dvfs}.
However, these governors are primarily optimized for bursty, intermittent, and CPU-bound workloads.
In contrast, LLM inference imposes a sustained load on the NPU, keeping it operating at maximum frequency by default.
% Hardware-based
Certain governors rely on hardware-exposed performance counters and state metrics~\cite{intelpstate}; however, most mobile NPUs do not offer such features.
Hence, NPU frequency scaling is often either static or manually-controlled~\cite{npu-dvfs}. 
% \bs{This contradicts "DVFS is supported for NPUs"}
Moreover, inference spans multiple components, whose frequencies must be co-managed to improve energy efficiency.
For example, CRAVE~\cite{crave} jointly tunes CPU, GPU, and memory frequencies on mobile platforms. However, it does not study different LLM inference workloads as they vary significantly in performance and efficiency, making offline profiles hard to generalize (\S\ref{sec:observation}).
% More related works on DVFS will be discussed in Section~\ref{sec:related}. 
% \bz{This paragraph is quite critical, but seems to be easily ignored in the background.}

\section{Characterizing On-device Inference}
\label{sec:observation}

% \TODO{update the figure.}

% While prior research has demonstrated that on-device inference can achieve impressive performance, its energy efficiency and thermal behavior have been largely overlooked.
% One study reports that mobile phones can process only around 500 prompts before their batteries are depleted~\cite{MELTing},
% significantly limiting the practical application.
% \bz{@Binqi: Could you please have a look at the observations and figures in this chapter? Is there a way to show more insights and connections between them and the design?}
To understand the energy and thermal behavior of on-device inference, 
we characterize throughput, power, and thermal metrics across LLMs on three platforms---a high-end phone, a laptop, and a development board~\cite{orangepi_5pro}. For certain measurements, we use external devices; details are provided in \S\ref{sec:evaluation}. 
%measured with an in-house component-level (e.g., NPU, Mem) power monitoring device for the phone/laptop, and a power meter for the board (\S\ref{sec:evaluation}).
Considering that the prefill stage simultaneously processes multiple tokens and is 15–25$\times$ faster than the decoding stage, the latter dominates the power consumption~\cite{MELTing} in most scenarios.
The prefill stage typically runs at the maximum hardware frequency to minimize TTFT.
This paper, therefore, explores the possibilities of energy savings only during decoding.
All test platforms primarily use the NPU for decoding.
Hence, for an LLM inference, \emph{we account only for the power consumed by the components active during decoding}---i.e., the NPU and memory (Fig.~\ref{fig:moti}).

% \smallskip

\begin{figure}[]
    \centering
    \includegraphics[width=0.95\columnwidth]{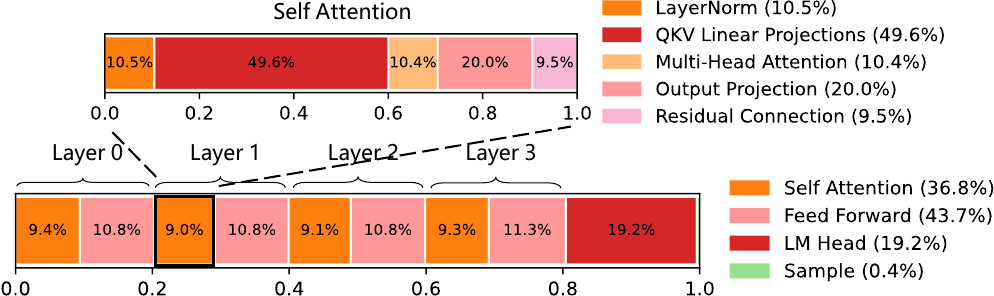}
    % \vspace{-2mm}
    \caption{Time breakdown of decoding a token in LLaMA2–1.3B.}
    \vspace{-2mm}
    \label{fig:breakdown-decode}
\end{figure}

\begin{figure}
    \centering
    \includegraphics[width=0.8\columnwidth]{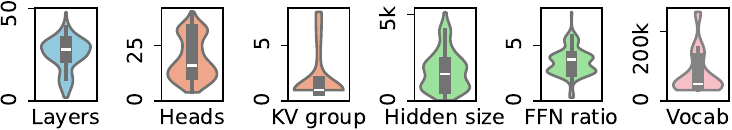}
    \caption{Hyperparameters of the top 126 LLMs ($\le$7B) from Hugging Face Open LLM Leaderboard~\cite{open-llm-leaderboard}. \textit{KV group} refers to the number of attention heads per KV head, i.e., $\frac{num_{attn\_head}}{num_{KV\_head}}$. \textit{FFN ratio} is equal to $\frac{intermediate\,size}{hidden\,size}$.}
    \label{fig:survey}
    \vspace{-2mm}
\end{figure}

% \smallskip
\textbf{Key hyperparameters.}
Figure~\ref{fig:breakdown-decode} shows the time breakdown for decoding a token using the 4-layer LLaMA2-1.3B. We can conclude that one decoding iteration is dominated by several matrix-multiplication operators, such as in the language modeling head (LM head), self-attention, and feed-forward networks (FFN). Moreover, the breakdown yields 6 key hyperparameters that determine counts and dimensions of all static matrix multiplications:
% The \emph{number of layers} determines the count of self-attention and FFN computations;
% the \emph{hidden (embedding) size}, \emph{number of attention heads}, and \emph{number of grouped key–value heads} shape matrix dimensions in the self-attention;
% the \emph{intermediate size} influences the FFN computations;
% and the \emph{vocabulary size} impacts the LM head.
\revise{
the \emph{number of layers}, \emph{hidden (embedding) size}, \emph{number of attention heads}, \emph{number of key-value heads}, \emph{intermediate size}, and \emph{vocabulary size}.
}

Additionally, we derive hyperparameter ranges from the top 126 popular $\le$7B LLMs on the Hugging Face \textit{Open LLM Leaderboard}~\cite{open-llm-leaderboard}, varying across different LLM families, like LLaMA~\cite{llama2-org}, Qwen~\cite{yang2024qwen2technicalreport}, Gemma~\cite{team2024gemma}, and Phi~\cite{javaheripi2023phi}. 
% Different LLM families exhibit only minor structural differences, while their backbones remain essentially the same. 
% ---such as different activation functions or alternative positional encoding schemes---
% As shown in Figure~\ref{fig:survey}, the number of layers mostly varies from 20 to 40 by scaling the size of each layer correspondingly. 
We depict the distribution of each hyperparameter in Figure~\ref{fig:survey}.
% Instead of \emph{number of grouped key–value heads}, 
\revise{
We use \textit{KV group} to denote the number of attention heads sharing each KV head, which is 1 in multi-head attention (MHA)~\cite{vaswani2023attentionneed} and greater than 1 in grouped query attention (GQA)~\cite{ainslie2023gqa}.
}
% as many LLMs still employ multi-head attention (MHA)~\cite{vaswani2023attentionneed}, rather than grouped query attention (GQA)~\cite{ainslie2023gqa}, the former of which has a single key-value head for each attention head. 
We use \emph{FFN ratio} to represent the ratio between \emph{intermediate size} and \emph{hidden size}, which is always 4 in some earlier LLMs, like GPT2~\cite{radford2019language}.
% Besides, unlike some earlier LLMs, such as GPT2~\cite{radford2019language}, the \emph{intermediate size} of which is always four times the hidden size, recent models have started to tune this ratio to improve performance and accuracy, as shown in the distribution of \textit{FFN ratio}.
We do not directly use them to achieve a broader range of parameter combinations;
because many share similar structures, which limits the diversity.
Instead, we generate 300 untrained models \footnote{Not fine-tuned. Used solely for throughput and power measurements.} by randomly selecting the six parameters within the ranges derived from the survey.

% \smallskip
\textbf{Evaluated LLMs.}
We evaluate LLMs suitable for on-device inference, including LLaMA2-1.3B/7B~\cite{llama2-org}\footnote{LLaMA2-1.3B refers to the modified version of LLaMA2~\cite{hfl2023chinesellama}.}, LLaMA3-3B~\cite{grattafiori2024llama3herdmodels}, Qwen2-1.5B~\cite{yang2024qwen2technicalreport}, and Gemma2-2B~\cite{team2024gemma}.
To generalize our observations to evolving LLMs, we also evaluate the 300 generated
untrained models.

We derive the following observations from experiments.

% \smallskip
\textit{\textbf{Observation 1: The energy efficiency of on-device inferences is highly sensitive to hardware configuration, where even modest throughput reductions can lead to substantial energy efficiency improvement.}}

\begin{figure}[]
    \centering
    \vspace{-2mm}
    \subfloat[Throughput.]{
        \includegraphics[width=0.42\columnwidth]{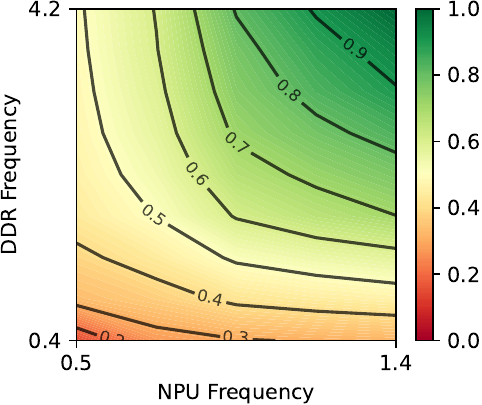}
    }
    \subfloat[Efficiency.]{
        \includegraphics[width=0.42\columnwidth]{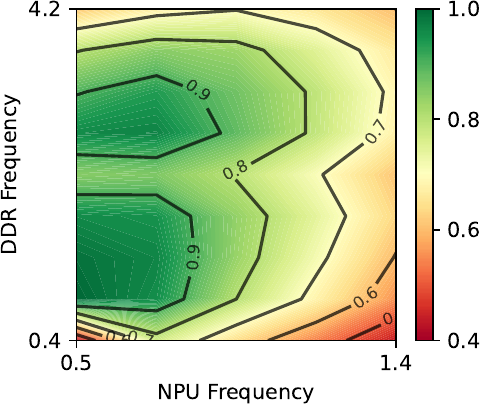}
    }
    \vspace{-2mm}
    % \caption{Decoding throughput and energy efficiency of different frequency settings of NPU and memory under Qwen2-1.5B Q4 on the phone. \textit{Def.}---default configuration at maximum frequencies; \textit{Opt.}---most energy-efficient configuration; \textit{Imp.}---configuration that improves efficiency by slightly reducing throughput. The arrows represent the Pareto front of throughput and efficiency.}
    \caption{Decoding throughput and energy efficiency of different frequency settings of NPU and DDR under Qwen2-1.5B (running solely) on phones, normalized to the maximum throughput and efficiency. Substantial headroom exists for improving efficiency.}
    \label{fig:energy-efficiency}
\end{figure}

% Setting the NPU and Mem to their maximum frequencies results in poor energy efficiency (\S\ref{sec:dvfs}).
% Figure~\ref{fig:energy-efficiency} presents the throughput and energy efficiency of Qwen2-1.5B with 4-bit quantization on the phone.
% Energy efficiency is defined as \emph{the number of tokens produced per joule}, considering only the hardware involved in an inference, i.e., the NPU and Mem.
% While the maximum frequencies yield the highest throughput (\textit{Def.}),
% the most energy-efficient setting (\textit{Opt.}) can improve efficiency by 62\%.
% Besides, slightly reducing throughput by 21\% can improve energy efficiency by 34\% (\textit{Imp.}).
% The efficiency table reveals non-monotonic trends with multiple local optima resulting from complex, non-linear relationships among frequency settings, the induced hardware utilization, and the resulting throughput and power.
% From the tables, we derive a Pareto front balancing throughput and energy efficiency. The arrows in Figure~\ref{fig:energy-efficiency} indicate the direction where energy efficiency is traded off for higher throughput.

Setting the NPU and memory to their maximum frequencies results in poor energy efficiency.
Figure~\ref{fig:energy-efficiency} presents the normalized throughput and energy efficiency of Qwen2-1.5B on the phone.
Energy efficiency is defined as \emph{the number of tokens produced per unit of energy}, considering only the hardware involved in inference, i.e., the NPU and DDR.
While the maximum frequencies (upper right corner in the figure) yield the highest throughput,
the energy-optimal setting can improve efficiency by 70\%.
Additionally, slightly reducing throughput by 20\% can improve energy efficiency  by 40\%.

{\centering\smallskip\fbox{\parbox{0.98\columnwidth}{
    \textbf{Insight 1:} \textit{Substantial headroom exists to improve energy efficiency by tuning hardware configurations while preserving the minimum QoE (leveraged by our design in \S{\ref{sec:design}}). 
    % $\to$ \textbf{Constrained optimization problem.}
    }
}}}
% \dr{do you think it makes sense to write at the end of this insight that ``(leveraged by our design in Sec. 4)''?}

% \textit{\textbf{Observation 2: The most energy-efficient configuration differs across models, so does the ranking of efficiency across configurations.}}
% \medskip
\smallskip
\textit{\textbf{Observation 2: The energy efficiency ranking, peak and scaled decoding throughput, and power consumption of the inference depend not only on the hardware configurations, but also on LLM model architectures.}}
% \smallskip

\begin{figure}[]
    \centering
    \includegraphics[width=0.95\columnwidth]{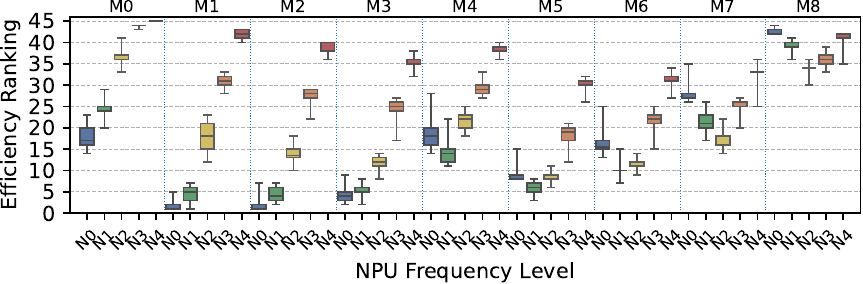}
    \vspace{-2mm}
    % \caption{Box plot of energy efficiency rankings (1st, 2nd, etc.) of hardware configurations \emph{across LLMs} on the phone.
    % Each box contains 300 LLMs.
    % Ranking varies significantly by model. The laptop and the board also show significant variation and are omitted. $Mx$ and $Nx$ denote the $x$th frequency level of Mem and NPU, where $x=0$ corresponds to the lowest available frequency level (the same applies below).}
    \caption{
    \revise{
    Box plot of energy efficiency rankings (1st, 2nd, etc.) of hardware configurations \emph{across LLMs} on the phone.
    Each box summarizes results from 300 LLMs.
    $Mx$ and $Nx$ denote the $x$-th frequency level of Mem and NPU, where $x=0$ corresponds to the lowest available frequency level (the same convention applies below).
    % Ranking varies significantly by model. 
    The laptop and the board exhibit similarly large variation. 
    }
    }
    \vspace{-4mm}
    \label{fig:ranking}
\end{figure}
\textbf{Efficiency ranking.}
Figure~\ref{fig:ranking} presents the rankings---in terms of energy-efficiency---of different configurations across LLMs on the phone.
% Notably, the most efficient frequency combination (i.e., $rank=1$) differs across models and platforms, and each configuration shows a wide range in its ranking.
Notably, the most efficient frequency setting (i.e., $rank=1$) generally occurs at lower levels (Mem and NPU)---e.g., $M1N0$---which is unlikely to meet the QoE target. 
Once the QoE is satisfied, and excluding a few inferior configurations (e.g., from $M4$ to $M8$, where lowering the NPU frequency consistently degrades both energy efficiency and throughput), models still differ in the configuration that leads to the highest efficiency. For each configuration, energy-efficiency rankings differ markedly across LLMs, as indicated by the wide spread in each box.
% In other words, for each configuration, the energy-efficiency rankings vary substantially across models, as reflected by the large spread in each box.
% \bz{rank 1 is not always the rank 1 of speed. More info could be derived.}
Considering that LLMs do not share a consistent monotonic efficiency trend, selecting a configuration to maximize the energy efficiency for an untested model is non-trivial.
% For instance, picking the lowest-frequency configuration that meets QoE often fails to deliver the highest efficiency, and no single traversal order (e.g., changing frequency from $MaNb$ to $Ma'Nb'$) can reliably yield the optimal QoE-compliant setting.

% {\centering\fbox{\parbox{0.98\columnwidth}{
%     \textbf{Insight 2:} \textit{The efficiency ranking is model-dependent and non-monotonic. Identifying the most efficient hardware configuration necessitates precise prediction of throughput and power.}
% }}}
\smallskip
{\centering\fbox{\parbox{0.98\columnwidth}{
    \textbf{Insight 2:} \textit{The efficiency ranking of hardware configurations is model-dependent and non-monotonic, which necessitates an accurate throughput and power prediction.
    % (addressed in \S{\ref{sec:selection}}).
    }
}}}
\smallskip

% \bz{The order here contradicts a bit with Sec.4, is it OK? (Insight 2 -> Sec.4.5)}

% \smallskip

% \textit{\textbf{Observation 3: Both the peak decoding throughput and the scaled decoding throughput under different hardware configurations vary across models.}}

\begin{figure}[]
    \includegraphics[width=0.95\columnwidth]{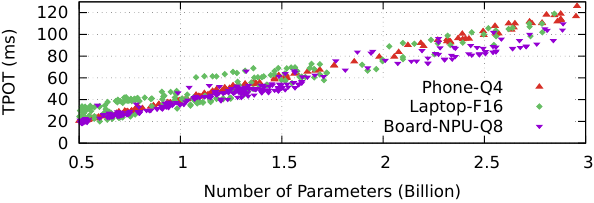}
    \vspace{-2mm}
    \caption{
    Time per output token (TPOT) of LLMs ($\ge$0.5B) with varying parameter counts, under the highest NPU and Mem frequencies. 
    % TPOT varies across LLMs, even with similar parameter counts.
    }
    \label{fig:est-bandwidth}
    \vspace{-2mm}
\end{figure}

\begin{figure}[]
    \centering
    \includegraphics[width=0.9\columnwidth]{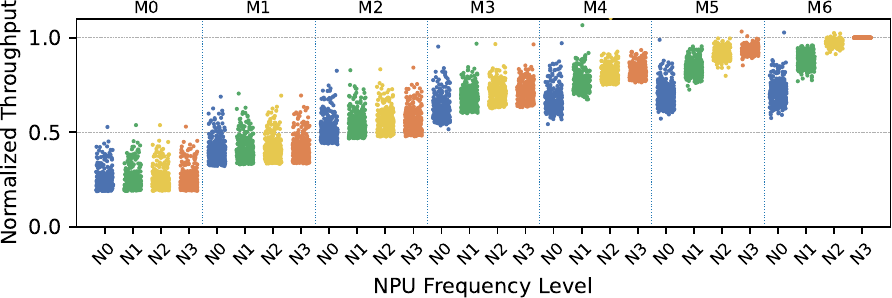}
    \vspace{-2mm}
    \caption{Strip plot of throughput normalized to the highest \emph{across LLMs} and configurations on the laptop.
    Each strip contains 300 LLMs.
    The phone/board show similar trends, thus omitted.
    % Throughput reduction relative to the peak varies significantly across models.
    }
    \vspace{-2mm}
    \label{fig:freq-deviation}
    % \Description{}
\end{figure}

\textbf{Peak and scaled decoding throughput.}
% A straightforward way to predict throughput is by parameter count, due to the memory-bounded nature of the decoding phase~\cite{pi2}.
\revise{
As the number of on-device LLMs grows, profiling their throughput across different architectures and hardware configurations becomes impractical.
Given the memory\-/bound nature of the decoding phase~\cite{pi2}, the model's parameter count serves as a coarse-grained indicator of throughput.
}
% However, we observe that models with similar sizes exhibit significantly different throughput.
Figure~\ref{fig:est-bandwidth} shows the time-per-output-token (TPOT, i.e., the reciprocal of throughput) of models with varying parameter counts across platforms at the highest frequency combination.
We observe that TPOT shows a quasi-linear relationship with parameter counts, whereas models of similar size still exhibit noticeable differences in TPOT, e.g., ranging from 34 ms to 62 ms for 1B models on the laptop.
% Even models with similar parameter counts,
% which are expected to perform identically due to the memory-bounded nature of the decoding phase~\cite{pi2},
% exhibit significant variations in throughput,
% e.g., ranging from 13-23 token/s for 1.3B models on the laptop.
% Notably, models of a similar size run at comparable speeds on the phone, likely due to saturation from the memory bandwidth limitation.
Moreover, scaling down the frequencies of NPU and Mem can also have distinct influences on different models. Figure~\ref{fig:freq-deviation} shows the throughput, normalized to peak performance, under various frequency combinations across models.
As shown, the throughput reduction ratio varies significantly with frequency choices, e.g., between 0.6 and 0.9 at ($M6N0$).
% Given the rapid evolution of LLMs---evidenced by Hugging Face recently surpassing 1 million models~\cite{hfmodels}---profiling each model offline individually is impractical in deployment.
% \bz{Can we move this to the intro?}
% \smallskip

{\centering\smallskip\fbox{\parbox{0.98\columnwidth}{
    \textbf{Insight 3:} \textit{Predicting the throughput of an LLM under different hardware configurations is essential, and throughput depends on the model's structure. 
    % Considering LLM's memory-boundedness, the number of model parameters helps in the prediction but with a limited accuracy (addressed in \S{\ref{sec:perf-pred}).
    \revise{
    Hence, the number of model parameters alone is insufficient for accurate prediction.
    }
    }
    % $\to$ \textbf{Throughput predictor.}
    }
}}

\begin{figure}[]
    \includegraphics[width=0.9\columnwidth]{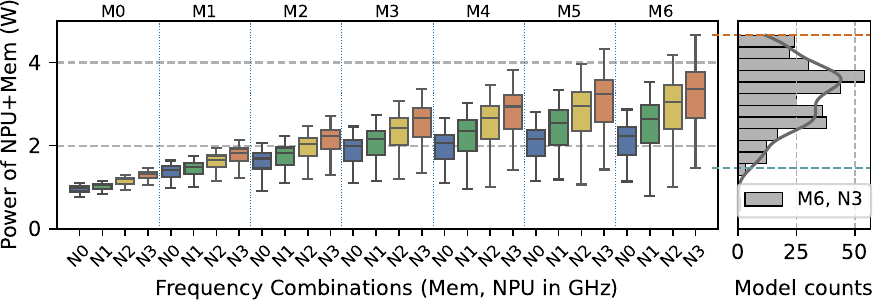}
    \vspace{-2mm}
    \caption{\textbf{Left: }Box plot of average power (NPU+Mem) \emph{across models} at different frequencies on the laptop, showing significant variation.
    Each box reflects \emph{inter-model variation}, not temporal fluctuation.
    \textbf{Right:} Power distribution \emph{across 300 models} at setting M6N3.
    Other platforms show similar trends and thus are omitted.}
    \vspace{-2mm}
    \label{fig:power-deviation}
\end{figure}

% \medskip
\smallskip
% \textit{\textbf{Observation 4: Power consumption of components varies across models, even at the same frequency.}}
\textbf{Power consumption.}
Due to the lack of component-level on-device power measurement, predicting the power consumption is essential for assessing the energy efficiency of unseen models.
Although frequency configurations determine fixed static power trends~\cite{eprof}, different LLMs still show notable variations in dynamic power even under the same settings.
Figure~\ref{fig:power-deviation} presents box plots of the average combined power consumption of the NPU and Mem during decoding under various frequency combinations.
For instance, even at a fixed setting---e.g., $M6N3$---power varies widely from 2.6W to 3.8W, with models spread evenly across the range.
This trend persists across platforms, primarily due to variations in NPU utilization and memory demands across models.

% {\centering\smallskip\fbox{\parbox{0.98\columnwidth}{
%     \textbf{Insight:} \textit{Power should be predicted based on hardware utilization, which varies across models (addressed in \S\ref{sec:power-pred}).}
% }}}
{\centering\smallskip\fbox{\parbox{0.98\columnwidth}{
    \textbf{Insight 4:} \textit{Power consumption should be predicted based on both hardware frequencies and utilization, where the latter depends on model architecture (addressed in \S\ref{sec:power-pred}).}
}}}

% \begin{figure}[]
%     \includegraphics[width=0.8\columnwidth]{./fig/power-vari-new.pdf}
%     \vspace{-2mm}
%     \caption{Power consumption in terms of NPU+Mem on the laptop. Only 30 randomly sampled LLMs from the 300 synthetic LLMs are illustrated. Each surface represents an LLM.}
%     \bzi{Is it a better plot than the box plot? Here we could motivate that we used the frequency as the input.}
%     \label{fig:power-deviation}
% \end{figure}

% \begin{figure}[]
%     \includegraphics[width=0.95\columnwidth]{./fig/co-run-thermal.pdf}
%     \vspace{-2mm}
%     \caption{Shell temperature of the phone under different scenarios. Scaling the frequency to get an acceptable throughput may significantly decelerate the temperature rise.}
%     \label{fig:co-run-thermal}
% \end{figure}

% \input{fig_tex/fig_thermal_obs}
% \medskip
\smallskip
% \textit{\textbf{Observation 3: Inference causes rapid heating, and hardware configurations chosen while ignoring the thermal dynamics may fail to preserve user‑comfort temperatures, especially under sustained workloads.}} 
\textit{\textbf{Observation 3: \revise{The output length (in tokens) is difficult to predict, and hardware configurations that ignore thermal dynamics may fail to preserve user-comfortable temperatures, especially under sustained workloads.}}} 

On-device LLM inference can lead to rapid heat generation due to sustained high load on the NPU, which arises from the auto-regressive decoding process where forward passes over the large model can continue for tens of seconds to several minutes especially in case of consecutive inference requests.
%manner and consecutive interactions. 
At the peak frequency setting, the shell temperature can exceed 42 ℃ in 100 s and 45 ℃ in 200 s, degrading the user experience.
Most existing systems throttle computing units only after a higher-temperature threshold is reached~\cite{aosp_overheat}, typically by lowering frequencies or pausing inference, which results in poor performance.

Even if the most energy-efficient configuration mitigates the temperature rise compared to operating at maximum frequency, the device may still exceed the thermal threshold over time. A better thermal management strategy must respond proactively when the shell temperature approaches the threshold. This necessitates predicting future shell temperatures under available hardware configurations while considering current and past states of the device. 

{\centering\smallskip\fbox{\parbox{0.98\columnwidth}{
    \textbf{Insight 5:} \textit{Uncontrolled temperature rise can lead to more severe degradation of user experience. A temperature-prediction model enables timely intervention and supports proactive thermal management (addressed in \S\ref{sec:thermal}).}
}}}

\section{Design of {\sys}}
\label{sec:design}

\subsection{Overview}

\begin{figure}[]
    \centering
    \includegraphics[width=\columnwidth]{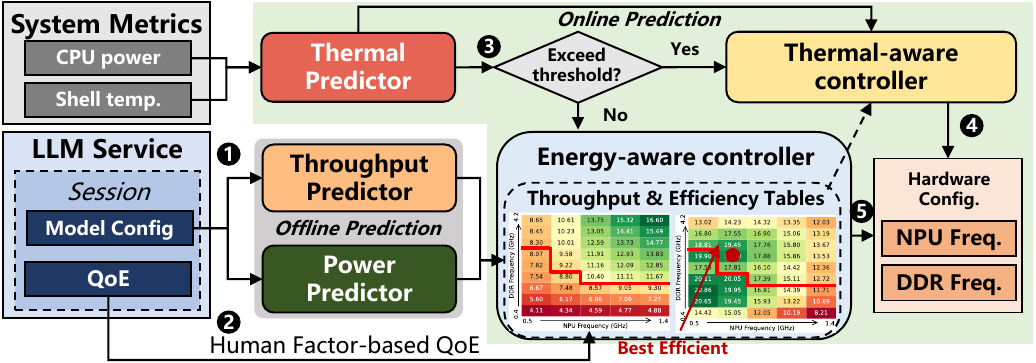}
    \vspace{-2mm}
    % \caption{Overview of {\sys}. {\sys} employs ML models to predict the throughput and power of unseen LLMs across hardware configurations, thereby maximizing energy efficiency while meeting the QoE requirement. Besides, {\sys} adopts a runtime thermal predictor to dynamically enable or disable a thermal-aware controller.
    % }
    \caption{Overview of {\sys}. ML models are employed to predict the throughput and power of unseen LLMs across hardware configurations to choose the most energy efficient one that meets the QoE requirement. 
    A runtime thermal predictor is adopted to dynamically enable or disable a thermal-aware controller.
    }
    \vspace{-2mm}
    \label{fig:overview}
\end{figure}

To enable energy-efficient on-device LLM inference, we propose {\sys}.
It maximizes energy efficiency while meeting QoE requirements, i.e., the decoding speed.
% including decoding speed and thermal constraints.
Figure~\ref{fig:overview} illustrates the overall design of {\sys}.

Given the structural information of the LLM model (i.e., hyperparameters),
{\sys} employs a machine-learning (ML) approach to accurately predict the throughput and power consumption of an unseen model across various hardware configurations, i.e., frequency combinations of NPU and Mem (\ding{182}).
The predictions are performed \emph{offline} during the model's initialization, and the results are stored in tables.
Leveraging the predicted tables,
{\sys} filters out choices that do not meet the throughput QoE requirement at runtime (\ding{183}, as lower bound).
\begin{comment}
To prevent overheating,
it further estimates the remaining response length based on statistical data and the number of generated tokens (\ding{184}).
Based on that, it predicts the shell temperature upon completion with respect to different hardware configurations (decoding speed).
% This \emph{online} prediction involves only simple linear regression and, hence, incurs negligible computational cost. \bz{We mentioned this in discussion. Could we delete it here?}
This step helps filter out choices that would lead to overheating (\ding{185}, as upper bound).
\end{comment}
To prevent overheating, {\sys} periodically uses a runtime thermal predictor to predict the shell temperature over a given time horizon (\ding{184}). If the predicted temperature exceeds the thermal threshold, {\sys} switches to a thermal-aware control mode in which a customized model predictive control (MPC) mechanism is applied (\ding{185}).
If not, it selects the most energy-efficient, QoE-satisfying hardware configuration (\ding{186}),
using a ranking-driven feedback mechanism.

\subsection{Throughput prediction}
\label{sec:perf-pred}

% \noindent
\textbf{White-box vs. black-box approaches.}
An intuitive approach to predict throughput is to use white-box analyses of a model's computational and memory operations.
However, the closed-source nature of most inference engines, operators, and NPU internals makes white-box modeling difficult.
Thus, {\sys} adopts an ML-based black-box approach
that takes the structural information of models as input (Insight 3 in \S\ref{sec:observation}).
This approach effectively predicts throughput by learning the computational and memory access patterns,
along with hardware characteristics, of the dominant General Matrix-Vector Multiplication (GEMV) operations---whose behavior is relatively predictable---during decoding.
% The black-box method relies entirely on collecting throughput data obtained through empirical measurements across varying hardware configurations and models,
% without needing details about accelerator internals.
The black-box method relies solely on empirically measured throughput across diverse hardware configurations and models, without requiring any accelerator-internal details.

\textbf{Monolithic vs. disaggregated models.}
As highlighted in Insight 3 in \S\ref{sec:observation}, the throughput of different hardware configurations should be predicted independently.
A straightforward approach is to use a monolithic model that jointly takes the structural information and the specific configuration---NPU and Mem frequencies---as input to predict throughput.
While intuitive, this method conflates model-dependent factors,
such as internal computation and memory access patterns,
with platform-dependent variations,
like the hardware's capability across frequency settings.
This coupling increases model complexity and
degrades accuracy.
% demands a larger training dataset to ensure accuracy.
Instead, {\sys} proposes disaggregated models: one that predicts the peak throughput at maximum frequencies and another that captures throughput reduction across configurations.
% Experiments (detailed setup below) show that the decoupling can reduce Mean Absolute Error (MAE) by 70\% compared to the monolithic model.
Experiments (see detailed setup below) show that this decoupling reduces Mean Absolute Percentage Error (MAPE) by up to 7.8\% compared to the monolithic model.

\textbf{Model selection.}
We train a throughput-prediction model using the collected throughput measurements of synthetic LLMs with varying hyperparameters and hardware configurations, see \S\ref{sec:observation}.
The nontrivial cost of generating and profiling models across configurations limits the dataset size (e.g., $<$1k samples), necessitating predictors that achieve high accuracy with limited data. 
Of the 300 synthetic models generated, 80\% are used for training with 5-fold cross-validation, and the remaining 20\% are reserved for testing.
We evaluate \emph{four} common ML-based regression models, each subjected to 1,000 iterations of random hyperparameter search.
% Table~\ref{tbl:regressor} compares the accuracy of different predictors under study,
Random forest (RF)~\cite{random-forest} achieves the best accuracy, with a MAPE of 3.27\%.
In contrast, a simple neural network---specifically a multilayer perceptron (MLP) \cite{mlp}---overfits due to the limited dataset and high input dimensionality, resulting in poor generalization (MAPE = 15.98\%). 
%With a limited number of data points and input dimensionality, a simple neural network---multilayer perceptron (MLP)---tends to overfit and show poor accuracy on unseen data~\cite{mlp} (MAPE=15.98\%). 
A polynomial (Poly) regressor~\cite{poly} fails to capture the complex non-linear interactions among the model, engine, and hardware (MAPE=33.89\%).
Although a support vector regressor (SVR)~\cite{svr} can model non-linear patterns via kernel functions, its accuracy remains suboptimal (MAPE=13.24\%) compared to RF.
Therefore, {\sys} employs RF for both predictors. The trained RF uses a maximum depth of 20 and no more than 200 trees. With computational complexity of $\mathcal{O}(TD)$, the prediction overhead is negligible relative to LLM model loading.
More sophisticated prediction techniques~\cite{edgetpu-predictor} could also be integrated to further improve accuracy, but such enhancements are orthogonal to our framework.

% \input{tab/ML_models}

% \noindent
\textbf{Input.}
%The peak throughput prediction model takes 6 hyperparameters---they determine the amount of computations and memory accesses in the decoding phase, see \S\ref{sec:observation}---as inputs, along with the total parameter count, which helps understand their internal relationships,
%predicting throughput under maximum frequencies.
To predict throughput at maximum frequencies, a \emph{peak-throughput predictor} is built that takes the following inputs: (i)~6 \emph{key hyperparameters} (see \S\ref{sec:observation}) governing computation and memory access during decoding, and (ii)~the total parameter count, which aids data augmentation.
% This model predicts throughput under the maximum frequency setting.
%The variation ratio predictor extends this by incorporating both NPU and Mem frequencies, predicting throughput scaling ratio relative to the maximum frequencies.
The \emph{scaled-throughput predictor} additionally incorporates NPU and memory frequencies as input and predicts throughput degradation relative to the peak value, and the baseline \emph{monolithic predictor} uses the same set of inputs.

\begin{figure}[]
    \vspace{-4mm}
    \subfloat[Throughput.]{
        \includegraphics[width=0.315\columnwidth]{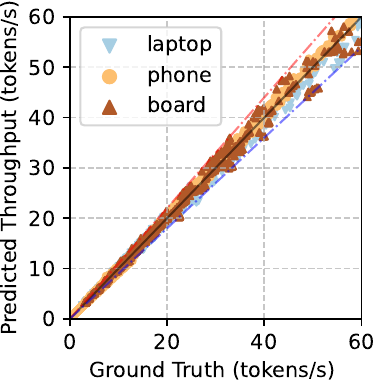}
        %\vspace{-1em}
        \label{fig:thpt-prediction}
    }
    \subfloat[Power.]{
        \includegraphics[width=0.32\columnwidth]{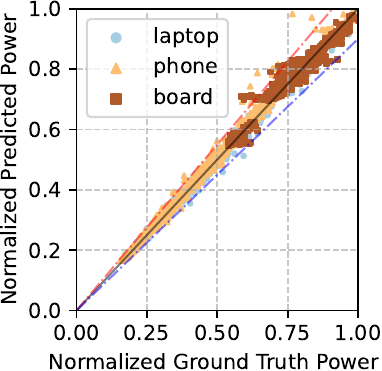}
        %\vspace{-1em}
        \label{fig:power-prediction}
    }
    \subfloat[Efficiency.]{
        \includegraphics[width=0.312\columnwidth]{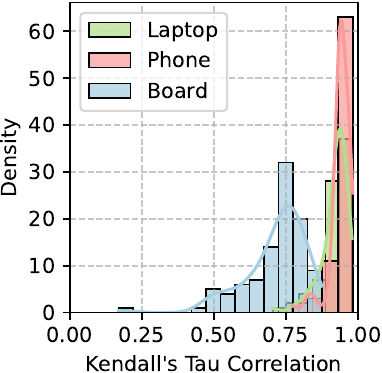}
        %\vspace{-1em}
        \label{fig:partial_order}
    }
    \vspace{-2mm}
    \caption{Accuracy of throughput and power prediction, the Kendall’s Tau correlation of predicted efficiency (the closer to 1, the better), showing high accuracy in predicting the efficiency ranking. The dotted line shows a 10\% error margin.}
    % \bzi{We need to add a separate power evaluation for NPU/Mem on laptop and phone.}
    \vspace{-0.5em}
\end{figure}

% \noindent
\textbf{Accuracy.}
% the peak throughput prediction,
% the scaled throughput ($\frac{\text{scaled throughput}}{\text{peak throughput}}$) prediction,
% and their combined prediction
Figure~\ref{fig:thpt-prediction} shows the prediction accuracy of throughput under specific frequency settings, obtained by combining the peak-throughput and scaling-factor predictions.
{\sys} accurately forecasts decoding throughput across different models, hardware configurations, and platforms.
The MAPEs of the combined throughput prediction are 4.2\%, 2.3\%, and 2.4\% on the laptop, phone, and board, respectively, whereas the best corresponding MAPEs of the \emph{monolithic predictor} are 12\%, 8.0\%, and 7.8\% when using the same set of regressors with comparable training effort.

{\centering\smallskip\fbox{\parbox{0.98\columnwidth}{
    \textbf{Design Decision:} \textit{Employ disaggregated ML models for throughput prediction: one predicting peak throughput and the other predicting relative reduction across configurations.
    % \dri{$\ldots$ throughput reduction specific to a hardware configuration}
    }
}}}

% Accuracy.
\subsection{Power prediction}
\label{sec:power-pred}
Since energy-efficiency rankings vary across LLMs (Insight~2 in \S\ref{sec:observation}),
and power consumption also differs by model (Insight~4 in \S\ref{sec:observation}),
accurate power prediction is key to identifying frequency settings that improve efficiency.

% \noindent
\textbf{White-box vs. black-box approaches.}
An intuitive approach to predicting power consumption is to model the underlying hardware.
However, as shown in prior work on server accelerators~\cite{npu-dvfs},
accurate modeling requires not only low-level implementation details but also fine-grained runtime measurements of power-related parameters---
\emph{measurements that are typically inaccessible on devices such as phones and laptops,
including detailed NPU and memory utilization.}
In contrast, ML-based methods are more suited for this setting~\cite{greathouse2018machine}.
% \bz{I think we need to highlight this.}
% Even though we observed that NPU and Mem loads vary across models, they are not always accessible on different platforms.
% \bohua{We could cite the paper that Debayan recommends \url{https://arxiv.org/pdf/2401.01826}}
Thus, {\sys} employs an ML-based black-box approach for power prediction,
similar to the throughput predictor (\S\ref{sec:perf-pred}).
% to predict the power consumption.
Using the same structural information---namely the six hyperparameters---and the NPU and Mem frequencies as input,
{\sys} employs an RF to predict the combined power consumption of the NPU and Mem (or the combined power consumption of the whole board when component-wise power measurements are unavailable).
%{\sys} employs an RF to predict the individual power consumption of the NPU and Mem (the combined power consumption of the board due to the lack of component-wise power measurements).
Figure~\ref{fig:power-prediction} shows the prediction accuracy across platforms,
with MAPE values of 1.5\%, 2.2\%, and 1.7\% for the laptop, phone, and board, respectively.

% \noindent
\textbf{Efficiency prediction.}
Along with throughput prediction, {\sys} computes energy efficiency---i.e., $\frac{\text{throughtput}}{\text{power}}$---across configurations and, more importantly, predicts their relative ranking.
This capability enables identifying the most efficient configuration among those satisfying both speed and thermal constraints.
Even when throughput prediction exhibits slight inaccuracies, {\sys} may still identify the most efficient configuration by relying on the predicted ordering and applying feedback control (\S\ref{sec:selection}).
% .\dr{reformulation suggested in comments}
%Due to slight inaccuracies in the throuhput prediction, {\sys} might fail to select the most efficient configuration but the efficiency loss in absolute terms might be insignificant. 
%Also, throughput QoE violation can be corrected via a feedback control (\S\ref{sec:selection}).
Figure~\ref{fig:partial_order} illustrates the Kendall’s Tau correlation~\cite{kendalltau} between the predicted and true energy-efficiency tables in the test set, highlighting the similarity in the \emph{ranking}.
% the energy efficiency table of the predicted values and the ground truth in the test set, highlighting the similarity in the \emph{ranking}.
As shown, for both phones and laptops, most correlations are close to 1, indicating that the partial order is largely preserved in the predictions.
Notably, the board shows a slightly weaker correlation, primarily because different configurations yield very similar efficiency values.
Thus, the energy loss from selecting a suboptimal configuration is negligible (evaluated in \S\ref{sec:eval-model}).

% {\centering\fbox{\parbox{0.98\columnwidth}{
% \textbf{Design Decision:} \textit{Adopt an ML model to predict power consumption with a high accuracy in terms of the ranking of energy efficiency across configurations.}
% }}}
{\centering\smallskip\fbox{\parbox{0.98\columnwidth}{
\textbf{Design Decision:} \textit{Adopt an ML model to predict power consumption, enabling accurate prediction of the energy-efficiency ranking across configurations.}
}}}

% \subsection{Output length estimation and \\thermal management}
\subsection{Thermal prediction and management}
\label{sec:thermal}

%Using the shell temperature at the end of an inference to filter out hardware configurations that may lead to overheating requires both a coarse estimation of output length and a prediction of the post-inference shell temperature, accounting for all heat-generating components (Observation 6 in \S\ref{sec:observation}).
\begin{comment}
By considering the shell temperature at the end of an inference, {\sys} can reject hardware configurations that may lead to overheating.
However, this requires a coarse estimate of the response length and, correspondingly, a prediction of the post-inference shell temperature while accounting for all heat-generating SoC components (Insight~5 in \S\ref{sec:observation}).
\end{comment} % MobiSys

{\sys} predicts whether the current frequency settings will push shell temperature past the threshold within a given time horizon, switching between energy‑aware and thermal‑aware modes as needed. Accordingly, it requires a prediction model accurate near the threshold and a thermal-aware controller capable of containing temperature rise within the same time window (Insight~5 in \S\ref{sec:observation})..

\begin{figure}
    \centering
    \includegraphics[width=0.48\columnwidth]{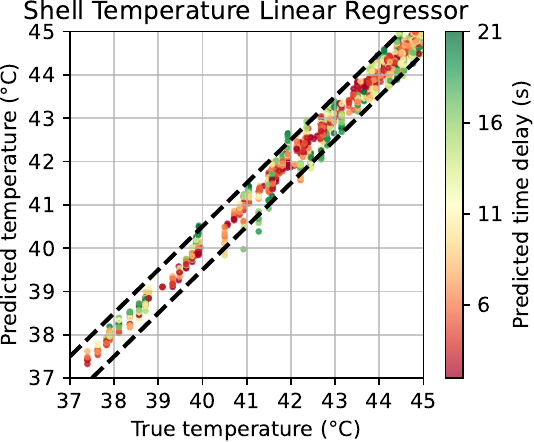}
    \caption{Thermal prediction accuracy in the test dataset. Dotted line: 0.5℃ error margin. It can accurately predict the temperature over the next 1-21 seconds.}
    % \caption{Thermal prediction accuracy on the test set, achieving accurate predictions up to 21 s. Dotted line: 0.5℃ error margin.}
    % \vspace{-4mm}
    \label{fig:thermal-pred}
\end{figure}

\textbf{Thermal prediction.}
% To predict the shell temperature upon completion (i.e., \textit{$PredictThermal$}), we observe that, in addition to the NPU, the CPU and GPU are the primary contributors to heat generation.
% However, due to a relatively constant GPU rendering load in typical LLM co-running scenarios (e.g., non-gaming apps),
% the shell temperature can be accurately predicted using only the power consumption by the CPU and the NPU.
% Incorporating additional components, including the GPU
% % \bz{Needs to be removed or replaced by another peripheral, maybe?}
% and Mem, does not lead to a significant improvement in prediction accuracy.
% Therefore, {\sys} uses a linear regression model to predict the shell temperature at  
% completion with input as follow:
The variation in shell temperature is governed by a highly complex and dynamic system. 
However, the thermal predictor adopted here only needs to operate within LLM-inference scenarios and over a constrained temperature range (e.g., 37-45℃), relying on recent temperature history and the current frequency settings. 
Our experiments show that Mem contributes little heat, whereas the NPU is the primary driver of temperature rise.
% Besides the NPU, parallel computations on the CPU and GPU can generate non-negligible heat (e.g., system services and rendering) and influence the evolution of the shell temperature during an inference.
% Concurrently with LLM inference, the GPU rendering load (e.g., from non-gaming apps) has been observed to remain relatively constant.
Accordingly, to predict shell temperature over a given time horizon, {\sys} incorporates predicted NPU power consumption together with recent temperature history to account for other system behaviors and environmental effects.
The predictor employs a linear regression model with the following inputs:
% (i)~historical data over an empirical 20-second window---specifically shell temperature and estimated CPU energy based on frequency and residency time;
(i)~an empirical 20-second history of shell temperature;
% and the estimated CPU energy, the latter is calculated based on the CPU frequency and its residency time;  
(ii)~the predicted NPU power consumption at the selected frequency; 
and (iii)~the target future time point. 
Using extensive experiments across different LLMs and settings at room temperature, we collect approximately 30k time segments.
Figure~\ref{fig:thermal-pred} shows the regression performance on the test dataset.
A simple linear regression model achieves high accuracy,
with an MAE of $0.16^\circ\mathrm{C}$ and $R^2$ of $0.988$ for predicting shell temperature 1–21 seconds into the future.

\textbf{MPC-based thermal-aware controller.}
Due to uncertainty in the future runtime of one or back-to-back LLM inferences and the dynamic nature of the system, MPC is well-suited for thermal management. We adopt a variant of MPC that assumes a constant control decision over the prediction horizon and employs an adaptive control period. The optimization objective is given in~\eqref{equ:obj}, where $u = (i_{npu}, i_{mem})$ denotes the NPU and Mem frequency settings, and $J_{N}$ represents the cost accumulated over $N$ steps.
\begin{align}\label{equ:obj}
    u^*=\arg\min_{u}J_{N}
\end{align}
One component of the cost function is the negative value of the tokens generated before the temperature threshold, encouraging the controller to maximize token production. However, it is unknown whether the current inference will complete within the next $N$ steps or whether another conversation will follow. 
%Thus, we consider a punishment for violating the thermal constraint by adding the number of tokens expected to be generated after the threshold is exceeding to the cost. 
To handle this uncertainty, we penalize thermal‑threshold violations by adding the expected number of tokens generated after the threshold is exceeded to the cost.
The resulting cost function is defined as:
\begin{equation}
\begin{aligned}
    J_{N}(u,\mathcal{T};\mathcal{R},\mathcal{P}_{npu},f)=\mathcal{R}[u](N\delta t - t^{\prime}) - \mathcal{R}[u]t^{\prime} &\\ 
    \text{s.t.} \quad f(\mathcal{T},\mathcal{P}_{npu}[u],t^{\prime})=T_{max}&
\end{aligned}
\end{equation}
where $\mathcal{T}$ denotes the recent shell‑temperature history, $\mathcal{R}$ and $\mathcal{P}_{npu}$ are the predicted throughput and NPU-power tables, $f$ is the thermal-prediction model, and $t^{\prime}$ is the time at which the temperature reaches the threshold. If the threshold is not reached within the prediction horizon, the cost reduces to
\begin{align}
    J_{N}(u,\mathcal{T};\mathcal{R},\mathcal{P}_{npu},f)=-\mathcal{R}[u]N\delta t
\end{align}

Our thermal-aware control strategy is outlined in Algorithm~\ref{alg:thermal_mpc}. The controller first collects the initial temperature history and sets the initial prediction period (lines 1-2). Upon entering the prediction loop, it resets the counter, retrieves the current configuration, and predicts the temperature over the specified horizon (lines 3-7). To reduce runtime overhead, we introduce a secondary thermal threshold, $T_{pre}$, to control the prediction period. If the temperature is estimated to exceed the thermal threshold, the thermal-aware controller is activated (lines 8-14); otherwise, the energy-aware controller remains active. At each step, the temperature history is updated by appending the current measurement and removing the oldest entry (lines 15-16).
{%
\setlength{\textfloatsep}{5pt} 
\SetAlCapFnt{\footnotesize}
\SetAlCapNameFnt{\footnotesize}
\SetAlFnt{\footnotesize}
\begin{algorithm}[t]
\revise{
    \SetKwInput{KwPar}{Parameter}
    \caption{MPC-based thermal-aware controller} % Closed-loop control flow
    \label{alg:thermal_mpc}
    \KwPar{
    $\delta t$: Time step for updating the temperature;\newline
    $N$: Prediction horizon;
    % $N_w$: Historical window length;
    }
    \KwIn{
    $\mathcal{R} \in \mathbb{R}^{m \times n}$: Predicted throughput table;\newline
    $\mathcal{P}_{npu} \in \mathbb{R}^{m \times n}$: Predicted NPU-power table;\newline
    $f(\mathcal{T}_w,P_{npu}, t)$: Thermal prediction model;\newline
    $r_{QoE}$: Throughput QoE requirement;\newline
    $T_{max}$: Thermal threshold;\newline
    $T_{pre}$: Temperature to trigger faster control;\newline
    $S$: Shared flag to indicate thermal-aware control;
    }
    % $\mathcal{T}\gets \{T_{-(N_w-1)\delta t},\ldots,T_{0}\}$\tcp*{Init temp. history}
    $\mathcal{T} \gets$ Initial 20-second temperature window;\\
    $(N_{step}, n) \gets (N, N)$\tcp*{Initial control period}
    % $n \gets N_{step}$\\
    \While{Inference service is running}{
        \If{$n \ge N_{step}$}{
            $n\gets0$;\\
            $u^{cur} \gets $ Current Mem and NPU frequency levels;\\
            $T_{N} \gets f(\mathcal{T}, \mathcal{P}_{NPU}[u^{cur}], N\delta t)$\tcp*{Prediction}
            \If{$T_{N} \ge T_{max}$}{
                $(N_{step}, S) \gets (1, true)$\tcp*{Thermal control}
                $u^*\gets\arg\min_{u} J(\mathcal{T},u;\mathcal{R},\mathcal{P}_{npu}, f, r_{QoE})$;
            }
            \uElseIf{$T_{N} \ge T_{pre}$}{
                $(N_{step}, S) \gets (1, false)$\tcp*{Period reduced}
            }
            \Else{
                $(N_{step}, S) \gets (N, false)$;
            }
            % \If{$S$}{
            %     $u^*\gets\arg\min_{u} J(\mathcal{T},u;\mathcal{R},\mathcal{P}_{npu}, f, r_{QoE})$
            % }
        }
        $\mathcal{T}.update(get\_temperature())$\tcp*{Update temp.}
        $n\gets n+1$; $sleep(\delta t)$;
    }
    \vspace{-2mm}
}
\end{algorithm}

}%

\begin{comment}
{\centering\smallskip\fbox{\parbox{0.98\columnwidth}{
    \textbf{Design Decision:} \textit{Adopt probabilistic output length estimation based on statistical data and apply linear regression for temperature prediction to manage thermal QoE at runtime.}
}}}
\end{comment}
{\centering\smallskip\fbox{\parbox{0.98\columnwidth}{
    \textbf{Design Decision:} \textit{Use linear regression for thermal prediction and an event-triggered, MPC-based thermal-aware controller.}
}}}

\subsection{Ranking-driven frequency selection}
\label{sec:selection}

{\sys} selects NPU and Mem frequencies at runtime to satisfy the QoE requirement. To reduce the search overhead, we derive a 2D Pareto frontier over energy efficiency and throughput and sort the configurations from most to least energy-efficient. 
The resulting ordered set is defined as:
\begin{equation}
\small
\begin{aligned}
    \mathcal{U}_{\Pi} &= \operatorname{argsort}_u\bigg( \\ &\Big\{\frac{\mathcal{R}(u)}{P(u)}\Big|u \in \mathcal{U}, \nexists u^{\prime} \in \mathcal{U} \colon \big(\mathcal{R}(u^{\prime}) > \mathcal{R}(u) \land \frac{\mathcal{R}(u^{\prime})}{P(u^{\prime})}>\frac{\mathcal{R}(u)}{P(u)}\big) \Big\}\bigg)
\end{aligned}
\end{equation}

\noindent
where $\mathcal{U}$ is the global decision space of $u$. The predicted throughput along the Pareto frontier is then
%
% \vspace{-2mm}
\begin{align}
    \mathcal{R}_{\Pi} = \left[\mathcal{R}[u] \mid u \in \mathcal{U}_{\Pi}\right]
\end{align}

Algorithm~\ref{alg:overall} outlines our ranking-driven, energy-aware frequency-selection mechanism integrated with the inference-service workflow. 
%Both sorted throughput values and configurations at the Pareto front are prepared during model initialization using the prediction models. 
During model initialization, the sorted throughput values and the Pareto‑frontier configurations are precomputed using the prediction models.
The most energy‑efficient configuration that satisfies the QoE requirement is then selected (line 1). 
During prefill, the peak frequencies are set (lines 2). Once decoding begins, the selected configuration is applied (line 3). During a decoding iteration, if the speed falls below the QoE requirement, the algorithm selects the next candidate in the sorted list (lines 4-11). Once the system enters thermal-aware mode ($S=\text{false}$), it no longer needs to choose hardware configurations.

{
\footnotesize
\setlength{\textfloatsep}{5pt} 
\SetAlCapFnt{\footnotesize}
\SetAlCapNameFnt{\footnotesize}
\SetAlFnt{\footnotesize}
\SetKwInput{KwPar}{Parameter}
\begin{algorithm}[t]
    \caption{Energy-aware controller.}
    \label{alg:overall}
    \KwIn{
    $r_{QoE}$: Throughput QoE requirement;\newline
    $\mathcal{R}_{\Pi}$: Sorted predicted throughput at the Pareto front;\newline
    % $\mathcal{E}_{\Pi}$: Predicted energy efficiency at the Pareto front, $\big\{\frac{T_F}{P_F}|F \in \mathcal{F} \land \nexists F^{\prime} \in \mathcal{F} \colon T_{F^{\prime}} > T_F, \frac{T_{F^{\prime}}}{P_{F^{\prime}}}>\frac{T_F}{P_F} \big\}$;
    $\mathcal{U}_{\Pi}$: Sorted frequency settings at the Pareto front;\newline
    $S$: Shared flag to indicate thermal-aware control;
    }
    $i^*\gets\arg\min_{i}\{\mathcal{R}_{\Pi}[i]>r_{QoE}\}$\\
    $u \gets $ Max. NPU and Mem frequency levels\tcp*{Peak for prefill.}
    % $\ldots$\tcp*{Prefill}
    $u \gets \mathcal{U}_{\Pi}[i^*]$ \tcp*{Decoding starts}
    $t_{old}\gets$ Current time stamp;\\
    \While{LLM generates a new token}{
        $t_{now}\gets$Current time stamp;\\
        $t_{old} \gets t_{now}$\\
        $r\gets \frac{1}{t_{now}-t_{old}}$\\
        \If{not S and $r < r_{QoE}$}{
                $i^*\gets i^*+1$;\\
                $u \gets \mathcal{U}_{\Pi}[i^*]$\tcp*{Feedback control}
        }
    }
    % \vspace{-2mm}    
\end{algorithm}
}

\begin{comment}
{\centering\smallskip\fbox{\parbox{0.98\columnwidth}{
    \textbf{Design Decision:} \textit{Periodical hardware configuration to improve energy efficiency while meeting the QoE constraints.}
}}}
\end{comment}
{\centering\smallskip\fbox{\parbox{0.98\columnwidth}{
    \textbf{Design Decision:} \textit{Adopt a ranking‑driven frequency‑selection strategy based on the 2D Pareto frontier, complemented by a feedback mechanism and coordination with the thermal‑aware controller.}
}}}

\subsection{Discussion}
\label{sec:discussion}

% \noindent
\textbf{Runtime overhead.}
Throughput and power predictions in {\sys} are performed during model initialization,
so runtime overhead is limited to table lookups and shell temperature prediction.
The latter is negligible, relying only on a lightweight linear regression with adaptive prediction and decision periods.
% Besides, this procedure occurs periodically (e.g., once per second), further mitigating its runtime overhead.
% Besides, it is performed periodically, and the period can be dynamically adjusted to reduce overheads, e.g., select longer periods when there are sufficient gaps for the shell temperature and the predicted throughput of the chosen configuration from their respective QoE requirements.

\textbf{Co-running LLM with other applications.}
Unlike LLM serving in the cloud, on-device inference tasks are likely to co-run with foreground apps, such as the voice assistant, note, and browser.
Whereas, we observe that daily-use apps have a \emph{negligible impact on both throughput and power} of inference tasks due to their lower utilization of the memory bandwidth and the NPU.
% their limited memory access bandwidth and minimal utilization of the NPU\dr{$\ldots$ their lower utilization of the memory bandwidth and the NPU}.
Table~\ref{tbl:mem-corun} shows the impact on throughput and power under several scenarios where the background inference co-runs with other apps.
We observe that there are minor impacts on throughput and NPU power, while the increase in the Mem power can be non-negligible in certain cases.
Thus, {\sys} adopts a fallback method.
When the memory bandwidth of the foreground app is negligible (the common case),
{\sys} uses Algorithm~\ref{alg:overall} to set frequencies as it would if the inference were running alone.
However, when \emph{the foreground app is memory-intensive} (the corner case),
{\sys} resorts to a \emph{feedback-based strategy} to identify the minimal configuration that meets the QoE, as the power consumption of the inference task and foreground app cannot be easily distinguished.
% This is because it cannot compute energy efficiency without being able to distinguish the power consumption by the inference task from that of the foreground app.

% Moreover, with {\sys}, a new voter is introduced in the system for requesting the minimal frequency on behalf of the inference tasks (similar to PM QoS~\cite{linux_pm_qos_doc}).
% In most cases, where the foreground application requires relatively low memory access bandwidth, the higher Mem frequency that {\sys} votes for takes effect.
% Otherwise, the application's vote for a higher frequency inherently overrides {\sys}'s.

% \begin{figure}[]
%     \centering
%     \includegraphics[width=0.9\columnwidth]{./fig/co-run-thpt-power.pdf}
%     \vspace{2mm}
%     \caption{Throughput and power deviations when co-running with foreground apps compared to an isolated run.}
%     \label{fig:mem-corun}
% \end{figure}

\begin{table}[]
\caption{Normalized throughput and power when co-running with foreground apps, w.r.t. a standalone run.}
% , which is 1.}
% \vspace{-2mm}
\label{tbl:mem-corun}
\resizebox{0.96\columnwidth}{!}{
\begin{tabular}{l|cccccccccc}
\toprule
% & \rotatebox{60}{{\sf Note}} & \rotatebox{60}{{\sf Brow.}} & \rotatebox{60}{{\sf Video}} & \rotatebox{60}{{\sf Voice}}  & \rotatebox{60}{{\sf Blog}} & \rotatebox{60}{{\sf Blog}} & \rotatebox{60}{{\sf Doc}}  & \rotatebox{60}{{\sf Paint}} & \rotatebox{60}{{\sf eShop}} & \rotatebox{60}{{\sf IM}}   \\ 
\textbf{Apps} & \sf Note & \sf Brow. & \sf Video & \sf Voice & \sf Blog & \sf Doc & \sf Paint & \sf eShop & \sf IM \\
\midrule
\textbf{Thpt.} & 0.98 & 0.89    & 0.99  & 0.95 & 0.96 & 0.99 & 0.97  & 0.92  & 0.98 \\
\textbf{$P_{Mem}$}  & 1.06 & 1.32    & 1.12  & 1.14 & 1.16 & 1.09 & 1.09  & 1.16   & 1.13 \\
\textbf{$P_{NPU}$}  & 1.03 & 0.90    & 1.04  & 1.01 & 1.04 & 1.04 & 1.01  & 0.99 & 1.06 \\
\bottomrule
\end{tabular}
}
\vspace{-6mm}
\end{table}

% Mem freq: min freq.

% \noindent
\textbf{Integration with acceleration methods.}
Model quantization~\cite{pmlr-v202-xiao23c} is a widely used technique for reducing memory and computational requirements, thereby increasing decoding speed.
Since quantization levels do not linearly affect throughput, {\sys} trains dedicated prediction models for specific levels---e.g., Q4 for phones and Q8 for boards (\S\ref{sec:evaluation}).
% \bz{We adopted quantization for phone and board, we need to rephrase this sentense.}
Another popular approach to reducing memory usage is the Mixture of Experts (MoE)~\cite{gshard}.
{\sys} can be extended to support MoE, because each expert maintains a fixed size, resulting in consistent decoding times. Only the number of experts needs to be considered additionally.

% Some other methods exploit model sparsity by storing only a subset of parameters in memory~\cite{llmflash, pi2}.
% Its throughput can be unpredictable and fluctuates over time.
% Similarly, speculative decoding~\cite{pmlr-v202-leviathan23a} uses smaller models for speculative decoding, leading to fluctuating throughput.\dr{reformulation suggested for the last 3 sentences}
Certain methods~\cite{llmflash} consider exploiting model sparsity and computing over a varying subset of parameters in each decode iteration, thereby making the throughput and power prediction challenging.
Speculative decoding~\cite{pmlr-v202-leviathan23a} also leads to a similar scenario.
Thus, {\sys} does \emph{not} natively support them.
Whereas, {\sys} could potentially support them by incorporating a history-based hit rate for throughput and power prediction, which we leave as future work.

% \noindent
\textbf{Adapting to new platforms.}
{\sys} has been validated on phones, a laptop, and a development board.
% \bz{Should we include mid-tier phone here?}
% \TODO{Efforts of generating models.}
Adapting to a new platform only requires collecting throughput and power data for the generated models ($\sim$300) to train platform-specific prediction models.
For each model, data collection requires $\sim$5 minutes considering available frequency combinations, and it can be automated as well.
Hence, {\sys} can be tuned for a new platform within a day.

% \noindent
\textbf{Adapting to XPUs.}
Although {\sys} primarily targets NPUs, as discussed in \S\ref{sec:LLM}, it can be extended to other XPUs---such as GPUs or TPUs---that support DVFS,
% as they exhibit similar energy efficiency characteristics
by capturing their energy efficiency characteristics.
% \dr{``$\ldots$ by capturing their energy efficiency characteristics'' or delete this part of the sentence}.
We have validated the effectiveness of {\sys} on the same development board, considering a GPU backend\footnote{Use MLC LLM~\cite{mlc-llm} as the inference engine with OpenCL backend to run on ARM Mali-G610 GPU in Orange Pi 5 Pro.} for the LLM inferences.
After training on 300 synthetic models and validating on four unseen real models, {\sys} achieves average throughput and power prediction MAPEs of 10.90\% and 5.27\%, respectively, and Kendall's Tau of 0.78 in efficiency, while the energy efficiency is improved by up to 26.1\%. 
\section{Implementation}\label{sec:impl} % Implementation

We integrate {\sys} into (i)~an in-house LLM-serving framework (similar to Ollama~\cite{ollama}) for tests over the phones and the laptop and (ii)~RKLLM~\cite{rkllm} for evaluations on Orange Pi 5 Pro.
At model initialization,
{\sys} invokes platform-specific predictors to estimate power consumption and performance under all hardware configurations; accordingly, it populates the performance and efficiency tables.
%storing the results into performance and efficiency tables.
The runtime frequency scaling is implemented in the callback function for each token generation (e.g., \texttt{LLMResultCallback} of RKLLM).
During an LLM inference, {\sys} performs dynamic frequency control in two phases.
During prefill, i.e., the first token generation, it sets the NPU and memory to operate at their maximum frequencies, minimizing time-to-first-token.
\begin{comment}
During decoding, frequency is adjusted based on QoE requirements (Algorithm~\ref{alg:overall}) at intervals determined by a constant number of generated tokens (implemented, e.g., in \texttt{LLMResultCallback} of RKLLM~\cite{rkllm}).\dr{revise}
\end{comment}
During decoding, it first sets the frequency selected from the 2D Pareto frontier that meets the QoE requirement.
In each decoding iteration, it adapts the frequency only when the actual speed falls below the QoE target (Algorithm~\ref{alg:overall}) and the thermal-aware controller is disabled. 
Thermal management---in particular, periodic temperature monitoring and prediction---is always active during LLM inference and sets a flag indicating when the controller should switch from energy‑aware to thermal‑aware frequency scaling.
Based on the accuracy of our thermal prediction model, we set the time step $\delta t=1$ and the prediction horizon $N=21$ to activate thermal-aware control as early as possible.
% triggered at fixed time intervals (e.g., once per second with a timer) or after a set number of tokens 
% (e.g., adjusts inside \texttt{LLMResultCallback} in RKLLM~\cite{rkllm}).
% \bz{The "or" here could still be a bit confusing. Maybe the latter only adapts to the phone for thermal aware control.}

To apply {\sys}'s control decisions,
a memory frequency governor is implemented that considers frequency requests from multiple voting sources (e.g., via sysfs),
similar to PM QoS in Linux~\cite{linux_pm_qos_doc}.
{\sys} is added as a new voter and votes for a minimum memory frequency (similar to \texttt{scaling\_} \texttt{min\_freq} in the \texttt{cpufreq} subsystem).
When the voters issue conflicting requests,
the governor enforces the maximum among the requested minimum frequencies.
Thus, {\sys} prevails only when its request exceeds others (e.g., from foreground applications), which occurs in most cases (\S\ref{sec:discussion}); otherwise, it defers to preserve foreground performance.
Further, it directly sets the operating frequency of the NPU which is exclusively used for running inference tasks.

%We note that the CPU energy used in thermal prediction is calculated using the residency time readings from the \texttt{time\_in\_state} node~\cite{cpustat}.\dr{revisit}\bz{I removed the CPU part to predict, should we delete this?}
% The CPU energy used in thermal prediction 

% \bs{The implementation section is a bit too short. If more space is needed, we can consider moving some side results to an appendix, which is not counted in the page limit}

% \bs{If possible, we can try to report some details / detailed numbers of the implementation to make it more convincing}

\section{Evaluations}
\label{sec:evaluation}

%We evaluate {\sys} to answer the following questions.
Our evaluation addresses the following questions.
\begin{itemize}[leftmargin=*]
% \begin{itemize}
    \item Q1: Does {\sys} precisely predict the throughput and energy efficiency of unseen models?
    \item Q2: Does it consistently meet the decoding speed QoE targets across different models?
    \item Q3: By how much does it improve energy efficiency while ensuring QoE?
    \item Q4: How does its prediction accuracy influence the obtained energy efficiency with respect to the oracle?
    \item Q5: Is the thermal-aware controller effective in sustaining the back-shell temperature within acceptable limits for a longer duration? 
    \item Q6: How much energy does it save in real-world scenarios?
\end{itemize}

\textbf{Platforms.}
We test {\sys} on the same platforms as in \S\ref{sec:observation},
i.e., a high-end phone (H-phone), a laptop and 
a development board~\cite{orangepi_5pro}.
Further, we test on a mid-tier phone (M-phone).
% including a high-end phone~\cite{mate70pro}, a mid-tier phone~\cite{p80}, a laptop~\cite{hmpc}, and a development board (Orange Pi 5 Pro~\cite{orangepi_5pro}).
% The phones and the laptop use an in-house inference engine that leverages the NPU for inference tasks,
%The phones and the laptop use an in-house inference engine with NPU as backend,
%while the board employs the RKLLM~\cite{rkllm} using RKNPU.
Power consumption for the phones and the laptop is measured with an in-house power monitor that uses an ADC to sample the voltage drop across the current-limiting resistor in each hardware component.
%Power consumption of the board is measured by an external power meter~\cite{fnirsi_fnb58}, capturing the overall system power.
We attach a power meter~\cite{fnirsi_fnb58} to the board that measures the total system power. 
To compute the power consumed by an inference, we subtract the idle power consumption of the board.
%We obtain the additional power consumed by an inference by subtracting the baseline power measured when idle.

\textbf{Baselines.}
We select the \textit{Default} configuration to reflect the behavior of "on‑demand" governors, which drive the NPU and DDR to their maximum frequencies under the sustained high load of LLM inference.
To evaluate the energy efficiency, we introduce \textit{Oracle} to represent the configuration that yields the highest efficiency based on the measured data (i.e., not predicted). Besides, a deadline-driven method (\textit{Deadline}) is selected for a comparison, which selects the minimal configuration that just meets the speed QoE requirement without considering energy efficiency. 
To evaluate the effectiveness of the thermal management, we compare it with \textit{Default} and the energy-aware setting without thermal management (\textit{Ener}).
\subsection{Prediction accuracy}

\begin{figure}[]
    \includegraphics[width=\columnwidth]{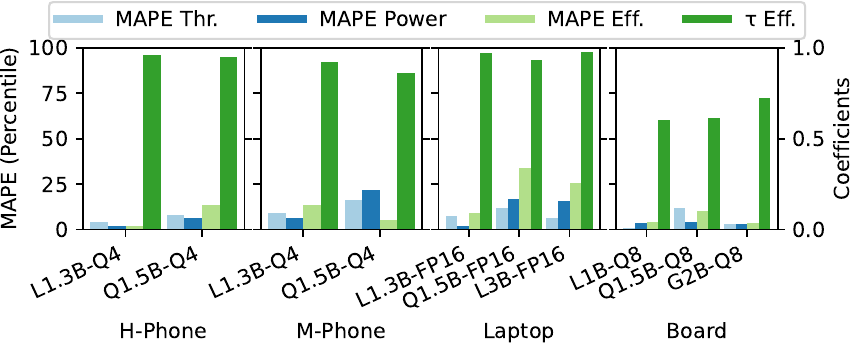}
    \vspace{-4mm}
    \caption{MAPE \emph{across frequencies} and Kendall's Tau ($\tau$, the closer to 1 the better) between predicted and ground truth in unseen real-world LLMs.
    G, L, and Q refer to Gemma2, LLaMA2/3.2, and Qwen2.}
    \vspace{-4mm}
    \label{fig:multi-cor}
\end{figure}

We evaluate the accuracy of {\sys}'s %throughput and power 
predictions for \textit{open-source}, \textit{production-ready} LLMs; they are \emph{not} part of the training dataset.
Figure~\ref{fig:multi-cor} reports the MAPE in throughput, power, and efficiency across different frequency combinations.
% Additionally, we show Kendall's Tau for the efficiency tables to show the accuracy in partial ordering, considering the efficiency of different configurations.\dr{reformulation suggested}
Also, using Kendall's Tau, we demonstrate the accuracy in the ranking of frequency combinations derived from the predicted energy efficiency values. 

% \noindent
%\textbf{Prediction Accuracy.}
%As shown in Figure~\ref{fig:multi-cor},
% \bz{We need to add the fullname of the models here as well}, 
We see that the throughput, power, and efficiency predictions exhibit high accuracy across platforms and models (\textbf{Q1}), with average deviations of 6.5\%, 12\%, 9.7\%, and 5.4\% for H-phone, M-phone, the laptop, and the board, respectively.
% \bz{Updated the results here. And for the average of each value, see in the comments.}
% Phone: throughput MAPE 7.7\%, power MAPE 3.9\%, efficiency MAPE: 7.7\%
% Laptop: throughput MAPE 9.7\%, power MAPE 10.5\%, efficiency MAPE: 22.7\%
% Board: throughput MAPE 6.37\%, power MAPE 3.44\%, efficiency MAPE: 6.43\%  
% More importantly, the Kendall’s Tau coefficients between the efficiency tables maintain an average of 0.94 across models on the phone and the laptop, indicating that {\sys} effectively identifies the most efficient configuration by using the partial order.
% % The boards exhibit lower coefficients due to minimal efficiency differences across configurations.
% The board exhibits lower coefficients, as efficiency varies minimally across configurations.
% However, the loss incurred from such inaccuracies compared to the oracle remains minor (\S\ref{sec:eval-model}).\dr{reformulation suggested}
\begin{comment}
More importantly, the average Kendall's Tau values are 0.95, 0.90, and 0.96 across models on H-phone, M-phone and the laptop, respectively. This shows high accuracy in determining the efficiency-based ranking of available configurations.
\end{comment}
More importantly, the average Kendall's Tau values are 0.95, 0.90, and 0.96 across models on the first three platforms, respectively. This shows high accuracy in determining the efficiency-based ranking of configurations.
A lower value (0.65) on the board is also acceptable considering that it leads to a minor loss in energy efficiency (\S\ref{sec:eval-model}), which is because the efficiency values vary marginally across configurations.

\subsection{Energy efficiency}
\label{sec:eval-model}

\begin{table*}[]
    \centering
    \caption{Efficiency relative increase w.r.t. \textit{Default} of {\sys} and \textit{Deadline} across models with QoE=5 or 10 tokens/s. \textit{Default} sets NPU and Mem to max frequency. \textit{Deadline} selects the frequencies that just fulfill QoE. Speed is measured in tokens/s;
    –: unable to meet QoE even at max frequency.}
    \vspace{-2mm}
    % \bzi{The speed of L1.3B on H-end phone does not match figure 4}
    \label{tab:eval-result}
    \resizebox{1.6\columnwidth}{!}{
    \begin{tabular}{cc|c|cc|cc|cc|cc}
        \toprule
        \multirow{2}{*}{Plat.} & \multirow{2}{*}{Model} & \textit{Default} & \multicolumn{2}{c|}{\textit{Deadline~(5t/s)}} & \multicolumn{2}{c|}{\textit{Deadline~(10t/s)}} & \multicolumn{2}{c|}{\sys~(5t/s)} & \multicolumn{2}{c}{\sys~(10t/s)} \\
        
            &   & Speed      & Speed  & \textit{Effi.} & Speed  & \textit{Effi.} & Speed  & \textit{Effi.} & Speed  & \textit{Effi.}\\
        \midrule
        % High-End
        H-end & L1.3B Q4 & 33.4 & 8.3 & 44\% & 11.7 & 65\% & 11.7 & \textbf{65\%} & 11.7 & \textbf{65\%}\\
        Phone & Q1.5B Q4  & 16.8 & 5.6 & 62\% & 10.1 & 50\% & 5.6 & \textbf{62\%} & 10.1 & \textbf{50\%} \\
        % & L1.3B FP16 & 11.5 & 6.1 & 5.3 & 9.3(51\%) & 10.2 & 7.2(17\%)\\
        % & Q1.5B FP16 & 7.3 & 4.3 & 5.6 & 5.7(31\%) & - & -\\
        \hline
        % Mid-tier
        M-tier & L1.3B Q4 & 28.2 & 7.8 & 4\% & 10.9 & 14\% & 16.7 & \textbf{26\%} & 16.7 & \textbf{26\%} \\
        Phone & Q1.5B Q4  & 14.8 & 5.9 & 16\% & 10.0 & 24\% & 9.2  & \textbf{27\%} & 11.0 & \textbf{26\%} \\
        % & L1.3B FP16 & 8.9 &  6.0 & 5.4 & 7.2(20\%) & - & -\\
        % & Q1.5B FP16 & 5.5 &  4.2 & 5.3 & 4.6(12\%) & - & -\\
        \hline
        % Laptop
        \multirow{3}{*}{Laptop} 
            & L1.3B  FP16 & 27.5 & 5.4 & -12\% & 12.6 & 3\% & 23.2 & \textbf{12\%} & 23.2 & \textbf{12\%} \\
            & Q1.5B FP16  & 18.5 & 6.2 & -16\% & 10.6 & 0\% & 15.7 & \textbf{10\%} & 15.7 & \textbf{10\%} \\
            & L3B FP16   & 8.8  & 5.5 & 5\% & - & - & 7.7  & \textbf{15\%} & -  & -\\
        \hline
        % Board
        \multirow{3}{*}{Board}
            & L1B Q8 & 21.5 & 6.2 & -3\% & 11.7 & -7\% & 6.3 & \textbf{14\%} & 20.8 & \textbf{8.6\%} \\
            % & Q1.5B Q8 & 14.6 & 4.1 & 11.9 & 4.5(8.9\%) & 11.9 & 4.5(8.9\%) \\
            & Q1.5B Q8 & 15.1 & 8.1 & -8.9\% & 10.1 & -2\% & 5.0 & \textbf{24\%} & 13.4 & \textbf{8.9\%} \\
            & G2B Q8 & 9.3 & 5.0 & -4\% & - & - & 8.1 & \textbf{11\%} & - & -\\
        \bottomrule
        
    \end{tabular}
    }
\end{table*}

\begin{figure}[]
    \centering
    % \vspace{-2mm}
    \begin{minipage}{\columnwidth}
    \centering
    \includegraphics[width=0.9\columnwidth]{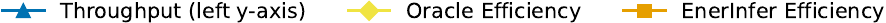}\\[-8pt]
    \subfloat[Q1.5B Q4 on high-end phone.]{
        \includegraphics[width=0.45\columnwidth]{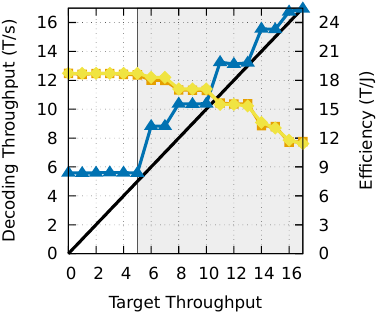}
        \label{fig:N-qwen2-1.5b-eval}
    }
    \subfloat[Q1.5B Q4 on mid-tier phone.]{
        \includegraphics[width=0.45\columnwidth]{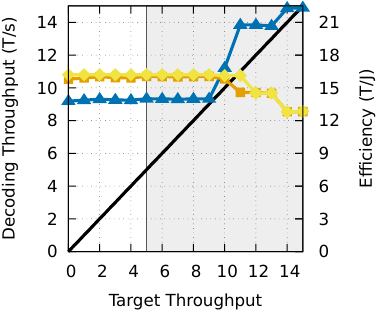}
        \label{fig:Xian-qwen2-1.5b-eval}
    }\\[-4pt]
    
    \subfloat[Q1.5B FP16 on the laptop.]{
        \includegraphics[width=0.45\columnwidth]{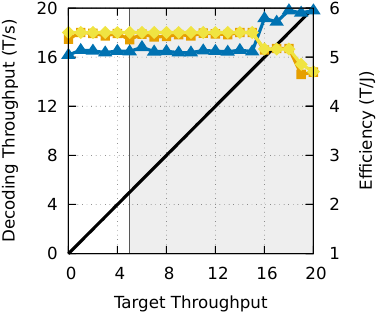}
        \label{fig:CPro-llama3-3b-eval}
    }
    \subfloat[Q1.5B Q8 on the board.]{
        \includegraphics[width=0.45\columnwidth]{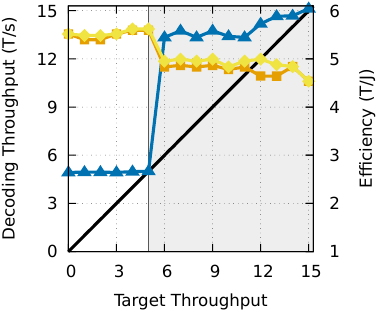}
        \label{fig:RKLLM-qwen2-1.5b-eval}
    }
    \vspace{-2mm}
    \caption{Actual efficiency and throughput of {\sys} across speed targets, using predicted results, compared to an oracle with ground-truth measurements. Shaded regions mark a practical QoE $>$ 5 tokens/s. {\sys} closely matches oracle across QoE targets.}
    \label{fig:eval}
    \end{minipage}
    % \vspace{-0.5em}
    \vspace{-2mm}
\end{figure}

Figure~\ref{fig:eval} illustrates the throughput and energy efficiency achieved by {\sys} under different QoE (decoding throughput) targets across selected LLMs and platforms.
% If QoE cannot be met even at the highest frequency, {\sys} uses it as a best-effort strategy.
The x-axis and the left y-axis represent target and achieved throughput, respectively.
Any points on and above the $y = x$ line meet the target.
The right y-axis represents the energy efficiency during the decoding phase.
\emph{Oracle} serves as the upper bound. %for efficiency under the QoE target.
%\textit{Efficiency} represents the energy efficiency of the {\sys}-selected configuration.
%\textit{Oracle} represents the configuration that yields the highest efficiency based on the measured data (i.e., not predicted), thus serving as the upper bound for efficiency under the QoE target.
Table~\ref{tab:eval-result} reports results for a wide range of models.

% \noindent
\smallskip
\textbf{QoE compliance.}
As shown in the figure, {\sys} consistently meets QoE targets---the achieved throughput is above the $y=x$ line---across all platforms and models (\textbf{Q2}).
When a relatively lower target is set, e.g., 4 token/s on phones, it selects a frequency combination that achieves a significantly higher throughput than the target.
This is because the lower frequency combination is not energy-efficient (Observation 4 in \S\ref{sec:observation}).
When the target exceeds the achievable throughput of the most efficient configuration, {\sys} selects a different configuration that offers a higher throughput, above the $y=x$ line, indicating that it maximizes the efficiency while meeting the throughput target.
% \noindent
\textbf{Efficiency improvement.}
% {\sys} improves energy efficiency while meeting QoE targets.
As given in Table~\ref{tab:eval-result}, with QoE set to 5 tokens/s, {\sys} achieves 50–65\%, 26–27\%, 10–15\%, and 9–24\% higher efficiency compared to \emph{Default}---NPU and Mem are set at maximum frequency---on the high-end phone, the mid-tier phone, the laptop, and the board, respectively (\textbf{Q3}). \textit{Deadline} does not guarantee better energy efficiency and may even reduce it.

\smallskip
\textbf{Comparison against the oracle.}
We examine the efficiency loss due to prediction inaccuracies by comparing {\sys} to the oracle. %that selects frequencies based on offline throughput and power measurements.
{\sys} achieves a near-oracle efficiency across models and platforms (\textbf{Q4}).
This tells that the predicted ranking of frequency combinations in terms of energy efficiency mostly follows the ground truth, 
even when the predicted efficiency values are slightly inaccurate (\S\ref{sec:power-pred}).
Hence, {\sys} can always find the most efficient, QoE-satisfying configuration by feedback adjustment.%\dr{revisit}
Even when the ranking is wrongly predicted, especially on the board, the efficiency loss remains minimal due to a lower efficiency variation across configurations.

\begin{comment}
We examine the efficiency loss due to inaccuracies in throughput and power predictions by {\sys}.
{\sys} achieves a near-oracle efficiency across models and platforms (\textbf{Q4}). 
This is because the partial ordering of frequency combinations mostly follows the ground truth even while using the slightly inaccurate, predicted energy efficiency values (\S\ref{sec:power-pred}), hence, {\sys} can find the most-efficient QoE-satisfying configuration.
Even when the partial order is wrongly predicted, especially on the board, the efficiency loss remains minimal due to a lower efficiency variation across configurations.
For certain targets, we see that \textit{Efficiency} is better than \textit{Oracle} but this is only because of small fluctuations in measurements or system behavior across different runs.
\end{comment}

\subsection{Thermal-aware controller evaluation}\label{sec:evalthermal}
\begin{figure}
    \centering
    \includegraphics[width=0.95\columnwidth]{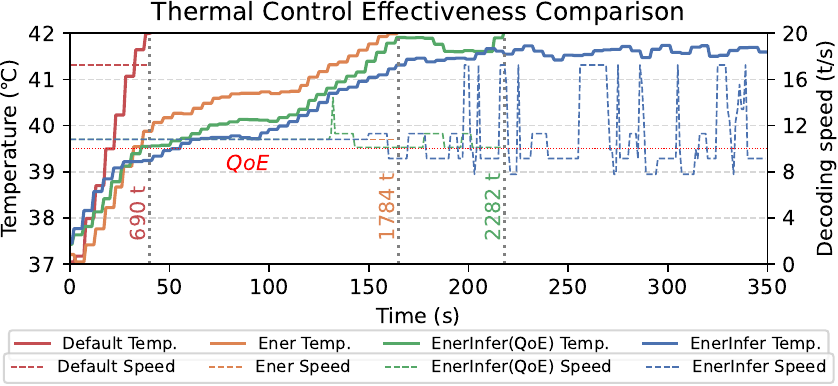}
    \caption{The shell temperature and decoding throughput under a back-to-back inference scenario before it reaches thermal threshold. \textit{Default}: default max. frequency setting. \textit{Ener}: energy-aware setting without thermal management. \textit{EnerInfer(QoE)}: our method with QoE constraint. {\sys}: our method without QoE constraint.
    }
    \vspace{-3mm}
    \label{fig:thermal_eval}
\end{figure}

To evaluate the effectiveness of our thermal-aware controller---which selects NPU and DDR frequencies---we design a stress test using back-to-back inferences with Qwen2.5-1.5B-Q4, starting from a shell temperature of 37 ℃. The results are illustrated in Figure~\ref{fig:thermal_eval}.
As shown, the \textit{Default} configuration has the highest throughput and the steepest temperature rise. It generates only 690 tokens in 40 seconds before reaching the threshold. As the most energy-efficient setting that meets the QoE, \textit{Ener} sustains a longer duration (165~s) and generates 1784 tokens. With our thermal management under the QoE constraint, {\sys} generates 27.9\% more tokens and lasts 32.1\% longer (\textbf{Q5}). When the QoE requirement is relaxed, {\sys} keeps the temperature below 42 ℃ for over 350~s while maintaining an acceptable average speed.

\subsection{Real-world deployment}

\begin{figure}[t]
    \centering
    \begin{minipage}{\columnwidth}
    \centering
    \includegraphics[width=\columnwidth]{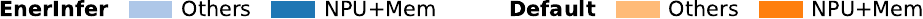}\\
    \subfloat[H-end phone.]{
        \includegraphics[width=0.25\columnwidth]{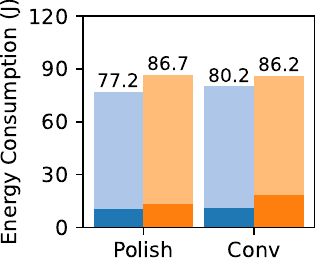}
        \label{fig:N-qwen2-1.5b-e2e}
    }
    \qquad
    \subfloat[M-tier phone.]{
        \includegraphics[width=0.25\columnwidth]{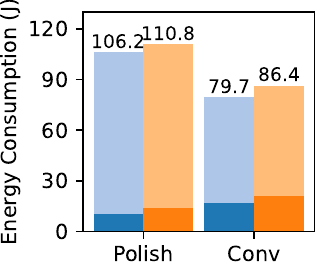}
        \label{fig:Xian-qwen2-1.5b-e2e}
    }
    \qquad
    \subfloat[Laptop.]{
        \includegraphics[width=0.25\columnwidth]{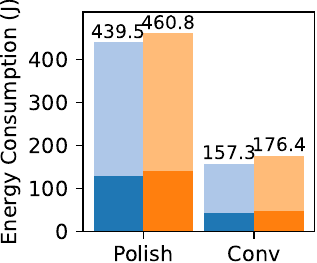}
        \label{fig:CPro-qwen2-1.5b-e2e}
    }
    % \vspace{-2mm}
    \caption{
    End-to-end total energy reduction by {\sys} in real-world scenarios. 
    Long ($\sim$50\%) post-inference display time dilutes the gains. \texttt{NPU+Mem} shows the inference energy.
    }
    \vspace{-4mm}
    \label{fig:realworld}
    \end{minipage}
    % \vspace{-2em}
\end{figure}

Figure~\ref{fig:realworld} reports the total energy reduction (including CPU, display, and other components) achieved by {\sys} in two typical on-device inference scenarios on the phones and the laptop.
% \bz{We could put the data of PC from our recent results.}
On the phones, the text polish task uses the Pangu-$\pi$ model~\cite{pangupi} and generates $\mathord{\sim}250$ tokens,
while the conversation task uses a Qwen2-1.5B Q4 model and generates $\mathord{\sim}200$ tokens.
On the laptop, both tasks use Qwen2-1.5B FP16 model and generate $\mathord{\sim}400$ and $\mathord{\sim}180$ tokens, respectively.
In both cases, LLM inference runs in the background while the screen actively renders foreground applications (a notes app and chat interface), adding additional load to CPU and GPU.
In the prefill phase, the NPU and Mem are configured to operate at their maximum frequencies, minimizing time-to-first-token.
During decoding, {\sys} enforces a QoE of 10 tokens/s to keep up with the display rate, allowing it to reduce energy while remaining imperceptible to users.
Moreover, with a fixed display rate, the end-to-end task duration remains constant with or without {\sys}, allowing a fair comparison of total energy consumption.

\textbf{End-to-end improvement.}
% 10.9%, 6.9%, 4.2%, 7.8%, 4.6%, 10.8%
Figure~\ref{fig:realworld} shows that {\sys} achieves energy savings between 4.2 -- 11\% compared to the \emph{Default}---defined in \S\ref{sec:evalthermal}---configuration (\textbf{Q6}). %of 11.0\%, 7.0\%, 4.2\%, 7.8\%, and 4.6\%, 10.8\% in the high-end phone, the mid-tier phone, and the laptop, respectively (\textbf{Q5}).
% The improvement is less pronounced compared to that when only the NPU and Mem energy consumption are considered.
% This is because the display, CPU, and GPU also consume part of the total energy in an end-to-end application.
This improvement is less pronounced than when considering only the NPU and Mem, because in an end-to-end application, the display, CPU, and GPU also draw part of the total energy.
% Besides, the decoding lasts only about half the time, and the remainder is spent on displaying the output. 
Further, decoding takes only about 50\% of the time, and the remainder is spent on prefill ($\mathord{\sim}$3\%) and displaying the output.
Thus, cutting energy use in decoding contributes moderately to the total end-to-end gain.
%Hence, the impact of reducing the energy consumption during decoding is diminished with respect to the end-to-end improvement.
% \bz{Show more convincing results. E.g. how many tokens/inferences can we run during one battery life.}

\section{Related Work}
\label{sec:related}

% \noindent

As discussed in \S\ref{sec:dvfs}, while prior work uses DVFS to optimize the energy efficiency of conventional workloads~\cite{cpu-dvfs, workload-dvfs, learning-dvfs, ddl-dvfs, crave}, such existing techniques are ineffective for LLM inferences.
A few works target energy‑efficient LLM serving in the cloud~\cite{kakolyris2024slo, greenerLLM, wordstowatts, atcGPUmiad, npu-dvfs, paul2025LLM-dvfs, Mauricio2024LLM-ener}.
% However, differences in workload characteristics (e.g., no batching on device), device capabilities (e.g., missing component-wise energy observability), and requirements (e.g., thermal limit) make these techniques less applicable for an on-device inference.
\revise{
However, differences in workload characteristics (e.g., no batching at the edge), hardware architectures (e.g., unified memory), and requirements (e.g., thermal limits) make these techniques less applicable to on-device inference.
}

% For mobile devices, Balasubramanian~et al.~\cite{balasubramanian2009energy} developed a protocol that reduces energy consumption of traditional mobile applications such as e-mail and web search. 
% MCDNN~\cite{han2016mcdnn} focuses on the scheduling and partitioning of DNN workloads between mobile devices and clouds under energy constraints. 
% LiKamWa et al.~\cite{likamwa2013energy} study the energy modeling and optimizations for continuous mobile vision workloads from CMOS image sensors. 
% AxoNN~\cite{dagli2022axonn} proposes an inference framework that schedules tasks across different hardware backends while considering a power budget, but it does not improve the energy efficiency.
\revise{
For mobile devices, Zhang~et al.~\cite{zhang2026rethinking} recently proposed a unified energy-aware governor spanning CPU, GPU, and memory for on-device LLM inference. 
%However, it lacks generalizability across device types and LLM architectures and does not account for thermal effects.
However, the approach depends on offline profiling—impractical on production devices and unable to generalize to unseen models—and assumes a device‑specific U‑shaped energy‑per‑token curve, limiting applicability across devices and LLM architectures.
}
MELTing~\cite{MELTing} and M4~\cite{yuan2024mobile} highlight the energy challenges of on-device LLM inference through benchmarking but do not provide solutions for energy savings. 
% \bz{All needs to be updated}
% yet they do not address these issues.

% \noindent
% \textbf{Black-box Prediction.}
Several works also use a black-box prediction. 
PETET~\cite{ni2022online} introduces an online throughput and power predictor based on the RF to identify a suitable backend for executing deep learning tasks.
It only considers fixed frequencies.
% using their hyperparameters as input.
Other works~\cite{yun2023, huang2010} estimate program execution time using methods like MLP or polynomial regression, which take syscall sequences or program features as input.
{\sys} employs a disaggregated model for predicting the throughput and power consumption of inference tasks under various configurations and uses it to guide the frequency selection.

% \noindent
% \textbf{Thermal management.}
Overheating is a common issue in mobile phones.
% zTT~\cite{kim2021ztt} learns system's thermal characteristics and employs DVFS on the CPU and GPU to prevent thermal throttling during general tasks.
A thermal\-/aware reinforcement learning is used in EdgeEngine~\cite{edgeengine} to optimize hardware frequencies for efficient edge inference.
DRLS~\cite{tan2024thermal} mitigates overheating during a deep neural network inference by deciding whether it should run on the GPU or NPU based on thermal prediction.
% TAPAS~\cite{jovan2025tapas} is the first work adopting power and thermal management for LLM inference, but only on the cloud.
\begin{comment}
{\sys} prevents overheating during LLM inference tasks by filtering out frequencies that may violate thermal limits.
\end{comment}
Due to the unpredictability in LLM output lengths, and the potential back-to-back inference requests, {\sys} adopts an MPC-based approach to keep the temperature under the threshold.
% lead to overheating.\dr{$\ldots$ may violate thermal limits}

\section{Conclusion and Future Work}\label{sec:conclusion}

On-device inference requires system-level optimization to improve throughput, energy efficiency, and thermal control.
%Emerging on-device inferences require system optimization to improve not only throughput but also energy efficiency and thermal management.
{\sys} enhances energy efficiency by accurately predicting the throughput and power consumption of an LLM across hardware configurations, and selecting the one that maximizes efficiency while preserving QoE, i.e., speed and thermal constraints.
The prediction models in {\sys} can be adapted to other AI workloads, and their outputs can assist in scheduling these workloads across heterogeneous on‑device backends to meet performance targets while improving overall system energy efficiency.
Further, we believe that more advanced control algorithms can be explored to optimize hardware-configuration selection as the device temperature approaches the threshold. 

\bibliographystyle{ACM-Reference-Format}
\balance
\bibliography{main}

@misc{pi2,
      title={PowerInfer-2: Fast Large Language Model Inference on a Smartphone}, 
      author={Zhenliang Xue and Yixin Song and Zeyu Mi and Xinrui Zheng and Yubin Xia and Haibo Chen},
      year={2024},
      eprint={2406.06282},
      archivePrefix={arXiv},
      primaryClass={cs.LG}, 
}

@misc{llmflash,
      title={LLM in a flash: Efficient Large Language Model Inference with Limited Memory}, 
      author={Keivan Alizadeh and Iman Mirzadeh and Dmitry Belenko and Karen Khatamifard and Minsik Cho and Carlo C Del Mundo and Mohammad Rastegari and Mehrdad Farajtabar},
      year={2024},
      eprint={2312.11514},
      archivePrefix={arXiv},
      primaryClass={cs.CL}, 
}

@inproceedings{DynamoLLM,
      title={DynamoLLM: Designing LLM Inference Clusters for Performance and Energy Efficiency}, 
     booktitle={IEEE International Symposium on High Performance Computer Architecture (HPCA)}, 
      author={Jovan Stojkovic and Chaojie Zhang and Íñigo Goiri and Josep Torrellas and Esha Choukse},
      year={2025},
}

@misc{greenerLLM,
      title={Towards Greener LLMs: Bringing Energy-Efficiency to the Forefront of LLM Inference}, 
      author={Jovan Stojkovic and Esha Choukse and Chaojie Zhang and Inigo Goiri and Josep Torrellas},
      year={2024},
      eprint={2403.20306},
      archivePrefix={arXiv},
      primaryClass={cs.AI}, 
}

@INPROCEEDINGS{wordstowatts,
  author={Samsi, Siddharth and Zhao, Dan and McDonald, Joseph and Li, Baolin and Michaleas, Adam and Jones, Michael and Bergeron, William and Kepner, Jeremy and Tiwari, Devesh and Gadepally, Vijay},
  booktitle={2023 IEEE High Performance Extreme Computing Conference (HPEC)}, 
  title={From Words to Watts: Benchmarking the Energy Costs of Large Language Model Inference}, 
  year={2023},
  volume={},
  number={},
  pages={1-9},
}

@inproceedings{aiops2024qiu,
  author  = {Qiu, Haoran and Mao, Weichao and Patke, Archit and Cui, Shengkun and Jha, Saurabh and Wang, Chen and Franke, Hubertus and Kalbarczyk, Zbigniew T. and Ba\c{s}ar, Tamer and Iyer, Ravishankar K.},
  title   = {Efficient Interactive LLM Serving with Proxy Model-based Sequence Length Prediction},
  year    = {2024},
  pages = {1--7},
  volume = {5},
  booktitle = {The 5th International Workshop on Cloud Intelligence / AIOps at ASPLOS 2024},
}

@inproceedings{pi,
author = {Song, Yixin and Mi, Zeyu and Xie, Haotong and Chen, Haibo},
title = {PowerInfer: Fast Large Language Model Serving with a Consumer-grade GPU},
year = {2024},
isbn = {9798400712517},
booktitle = {ACM SIGOPS 30th Symposium on Operating Systems Principles},
pages = {590–606},
numpages = {17},
series = {SOSP '24}
}

@misc{HeteroLLM,
      title={HeteroLLM: Accelerating Large Language Model Inference on Mobile SoCs platform with Heterogeneous AI Accelerators}, 
      author={Le Chen and Dahu Feng and Erhu Feng and Rong Zhao and Yingrui Wang and Yubin Xia and Haibo Chen and Pinjie Xu},
      year={2025},
      eprint={2501.14794},
      archivePrefix={arXiv},
      primaryClass={cs.DC}, 
}

@misc{pangupi,
      title={PanGu-$\pi$: Enhancing Language Model Architectures via Nonlinearity Compensation}, 
      author={Yunhe Wang and Hanting Chen and Yehui Tang and Tianyu Guo and Kai Han and Ying Nie and Xutao Wang and Hailin Hu and Zheyuan Bai and Yun Wang and Fangcheng Liu and Zhicheng Liu and Jianyuan Guo and Sinan Zeng and Yinchen Zhang and Qinghua Xu and Qun Liu and Jun Yao and Chao Xu and Dacheng Tao},
      year={2023},
      eprint={2312.17276},
      archivePrefix={arXiv},
      primaryClass={cs.CL}, 
}

@inproceedings{llmnpu,
author = {Xu, Daliang and Zhang, Hao and Yang, Liming and Liu, Ruiqi and Huang, Gang and Xu, Mengwei and Liu, Xuanzhe},
title = {Fast On-device LLM Inference with NPUs},
year = {2025},
booktitle = {ACM International Conference on Architectural Support for Programming Languages and Operating Systems},
pages = {445–462},
numpages = {18},
series = {ASPLOS '25}
}

@misc{personalllm,
      title={Personal LLM Agents: Insights and Survey about the Capability, Efficiency and Security}, 
      author={Yuanchun Li and Hao Wen and Weijun Wang and Xiangyu Li and Yizhen Yuan and Guohong Liu and Jiacheng Liu and Wenxing Xu and Xiang Wang and Yi Sun and Rui Kong and Yile Wang and Hanfei Geng and Jian Luan and Xuefeng Jin and Zilong Ye and Guanjing Xiong and Fan Zhang and Xiang Li and Mengwei Xu and Zhijun Li and Peng Li and Yang Liu and Ya-Qin Zhang and Yunxin Liu},
      year={2024},
      eprint={2401.05459},
      archivePrefix={arXiv},
      primaryClass={cs.HC}, 
}

@article{speakingspeed,
  title={British English-speaking speed 2020},
  author={Wang, Li},
  journal={Acad. J. Humanit. Soc. Sci},
  volume={4},
  pages={93--100},
  year={2021}
}

@article{readingspeed,
  title={Standardized assessment of reading performance: The new international reading speed texts IReST},
  author={Trauzettel-Klosinski, Susanne and Dietz, Klaus and IReST Study Group},
  journal={Investigative ophthalmology \& visual science},
  volume={53},
  number={9},
  pages={5452--5461},
  year={2012},
  publisher={The Association for Research in Vision and Ophthalmology}
}

@inproceedings{thermallimit,
  title={Ergonomic temperature limits for handheld electronic devices},
  author={Berhe, Mulugeta K},
  booktitle={International Electronic Packaging Technical Conference and Exhibition},
  volume={42789},
  pages={1041--1047},
  year={2007}
}

@inproceedings {atcGPUmiad,
author = {Haoran Qiu and Weichao Mao and Archit Patke and Shengkun Cui and Saurabh Jha and Chen Wang and Hubertus Franke and Zbigniew Kalbarczyk and Tamer Ba{\c s}ar and Ravishankar K. Iyer},
title = {Power-aware Deep Learning Model Serving with {$\mu$-Serve}},
booktitle = {2024 USENIX Annual Technical Conference (USENIX ATC 24)},
year = {2024},
isbn = {978-1-939133-41-0},
address = {Santa Clara, CA},
pages = {75--93},
publisher = {USENIX Association},
month = jul
}

@inproceedings{MELTing,
author = {Laskaridis, Stefanos and Katevas, Kleomenis and Minto, Lorenzo and Haddadi, Hamed},
title = {MELTing Point: Mobile Evaluation of Language Transformers},
year = {2024},
booktitle = {Proceedings of the 30th Annual International Conference on Mobile Computing and Networking},
pages = {890–907},
numpages = {18},
series = {MobiCom}
}

@misc{Gemini,
    title={Chat with Gemini to supercharge your creativity and productivity.},
    author={Google},
    year={2025},
    howpublished={\url{
https://store.google.com/intl/en/ideas/categories/ai/}},
    note  = {Accessed 7 Feb 2025}
}

@misc{apple-on-device,
    title={Introducing Apple’s On-Device and Server Foundation Models.},
    author={Apple},
    year={2024},
    howpublished={\url{https://machinelearning.apple.com/research/introducing-apple-foundation-models}},
    note  = {Accessed 7 Feb 2025}
}

@misc{Andes-QoE,
      title={Andes: Defining and Enhancing Quality-of-Experience in LLM-Based Text Streaming Services}, 
      author={Jiachen Liu and Jae-Won Chung and Zhiyu Wu and Fan Lai and Myungjin Lee and Mosharaf Chowdhury},
      year={2024},
      eprint={2404.16283},
      archivePrefix={arXiv},
      primaryClass={cs.DC}, 
}

@misc{hfmodels,
    title={Exponential growth brews 1 million AI models on Hugging Face},
    author={Benj Edwards},
    year={2024},
    howpublished={\url{https://arstechnica.com/information-technology/2024/09/ai-hosting-platform-surpasses-1-million-models-for-the-first-time/}},
    note  = {Accessed 7 Feb 2025}
}

@misc{llama2-org,
      title={Llama 2: Open Foundation and Fine-Tuned Chat Models}, 
      author={Hugo Touvron and Louis Martin and Kevin Stone and Peter Albert and Amjad Almahairi and Yasmine Babaei and Nikolay Bashlykov and Soumya Batra and Prajjwal Bhargava and Shruti Bhosale and Dan Bikel and Lukas Blecher and Cristian Canton Ferrer and Moya Chen and Guillem Cucurull and David Esiobu and Jude Fernandes and Jeremy Fu and Wenyin Fu and Brian Fuller and Cynthia Gao and Vedanuj Goswami and Naman Goyal and Anthony Hartshorn and Saghar Hosseini and Rui Hou and Hakan Inan and Marcin Kardas and Viktor Kerkez and Madian Khabsa and Isabel Kloumann and Artem Korenev and Punit Singh Koura and Marie-Anne Lachaux and Thibaut Lavril and Jenya Lee and Diana Liskovich and Yinghai Lu and Yuning Mao and Xavier Martinet and Todor Mihaylov and Pushkar Mishra and Igor Molybog and Yixin Nie and Andrew Poulton and Jeremy Reizenstein and Rashi Rungta and Kalyan Saladi and Alan Schelten and Ruan Silva and Eric Michael Smith and Ranjan Subramanian and Xiaoqing Ellen Tan and Binh Tang and Ross Taylor and Adina Williams and Jian Xiang Kuan and Puxin Xu and Zheng Yan and Iliyan Zarov and Yuchen Zhang and Angela Fan and Melanie Kambadur and Sharan Narang and Aurelien Rodriguez and Robert Stojnic and Sergey Edunov and Thomas Scialom},
      year={2023},
      eprint={2307.09288},
      archivePrefix={arXiv},
      primaryClass={cs.CL}, 
}

@misc{grattafiori2024llama3herdmodels,
      title={The Llama 3 Herd of Models}, 
      author={Llama Team, Meta},
      year={2024},
      eprint={2407.21783},
      archivePrefix={arXiv},
      primaryClass={cs.AI}, 
}

@misc{yang2024qwen2technicalreport,
      title={Qwen2 Technical Report}, 
      author={An Yang and Baosong Yang and Binyuan Hui and Bo Zheng and Bowen Yu and Chang Zhou and Chengpeng Li and Chengyuan Li and Dayiheng Liu and Fei Huang and Guanting Dong and Haoran Wei and Huan Lin and Jialong Tang and Jialin Wang and Jian Yang and Jianhong Tu and Jianwei Zhang and Jianxin Ma and Jianxin Yang and Jin Xu and Jingren Zhou and Jinze Bai and Jinzheng He and Junyang Lin and Kai Dang and Keming Lu and Keqin Chen and Kexin Yang and Mei Li and Mingfeng Xue and Na Ni and Pei Zhang and Peng Wang and Ru Peng and Rui Men and Ruize Gao and Runji Lin and Shijie Wang and Shuai Bai and Sinan Tan and Tianhang Zhu and Tianhao Li and Tianyu Liu and Wenbin Ge and Xiaodong Deng and Xiaohuan Zhou and Xingzhang Ren and Xinyu Zhang and Xipin Wei and Xuancheng Ren and Xuejing Liu and Yang Fan and Yang Yao and Yichang Zhang and Yu Wan and Yunfei Chu and Yuqiong Liu and Zeyu Cui and Zhenru Zhang and Zhifang Guo and Zhihao Fan},
      year={2024},
      eprint={2407.10671},
      archivePrefix={arXiv},
      primaryClass={cs.CL}, 
}

@inproceedings{npu-dvfs,
author = {Wang, Zibo and Zhang, Yijia and Wei, Fuchun and Wang, Bingqiang and Liu, Yanlin and Hu, Zhiheng and Zhang, Jingyi and Xu, Xiaoxin and He, Jian and Wang, Xiaoliang and Dou, Wanchun and Chen, Guihai and Tian, Chen},
title = {Using Analytical Performance/Power Model and Fine-Grained DVFS to Enhance AI Accelerator Energy Efficiency},
year = {2025},
isbn = {9798400706981},
booktitle = {ACM International Conference on Architectural Support for Programming Languages and Operating Systems},
pages = {1118–1132},
numpages = {15},
series = {ASPLOS '25}
}

@article{tan2024thermal,
  title={Thermal-aware scheduling for deep learning on mobile devices with NPU},
  author={Tan, Tianxiang and Cao, Guohong},
  journal={IEEE Transactions on Mobile Computing},
  year={2024},
  publisher={IEEE}
}

@inproceedings{yun2023,
  title={Runtime {WCET} Estimation Using Machine Learning},
  author={Yun, Sangwoon and Kang, Kyungtae},
  booktitle={Annual International Conference on Mobile Computing and Networking (MobiCom)},
  pages={1--3},
  year={2023}
}

@inproceedings{huang2010,
  author = {Huang, Ling and Jia, Jinzhu and Yu, Bin and Chun, Byung-Gon and Maniatis, Petros and Naik, Mayur},
  title = {Predicting execution time of computer programs using sparse polynomial regression},
  year = {2010},
  booktitle = {Advances in neural information processing systems (NeurIPS)},
  pages = {883-891},
}

@inproceedings{yuan2024mobile,
author = {Yuan, Jinliang and Yang, Chen and Cai, Dongqi and Wang, Shihe and Yuan, Xin and Zhang, Zeling and Li, Xiang and Zhang, Dingge and Mei, Hanzi and Jia, Xianqing and Wang, Shangguang and Xu, Mengwei},
title = {Mobile Foundation Model as Firmware},
year = {2024},
pages = {279–295},
numpages = {17},
booktitle={Annual International Conference on Mobile Computing and Networking (MobiCom)},
}

@inproceedings{ni2022online,
  title={Online performance and power prediction for edge TPU via comprehensive characterization},
  author={Ni, Yang and Kim, Yeseong and Rosing, Tajana and Imani, Mohsen},
  booktitle={2022 Design, Automation \& Test in Europe Conference \& Exhibition (DATE)},
  pages={612--615},
  year={2022},
  organization={IEEE}
}

@misc{fnirsi_fnb58,
  author       = {FNIRSI},
  title        = {FNB58 USB Fast Charge Tester},
  year         = 2025,
  howpublished = {\url{https://www.fnirsi.com/products/fnb58}},
  note         = {Accessed 25 Apr 2025},
}

@misc{orangepi_5pro,
  author       = {Orange Pi},
  title        = {Orange Pi 5 Pro},
  year         = 2025,
  howpublished = {\url{http://www.orangepi.org/}},
  note         = {Accessed 25 Apr 2025},
}

@misc{rkllm,
  author       = {Rockchip},
  title        = {{RKLLM Project}},
  year         = 2025,
  howpublished = {\url{https://github.com/airockchip/rknn-llm}},
  note         = {Accessed 25 Apr 2025},
}

@inproceedings{crave,
author = {Mukherjee, Dipayan and Hachem, Sam and Bao, Jeremy and Madsen, Curtis and Ma, Tian and Ghose, Saugata and Agha, Gul},
title = {CRAVE: Analyzing Cross-Resource Interaction to Improve Energy Efficiency in Systems-on-Chip},
year = {2025},
booktitle = {Proceedings of the Twentieth European Conference on Computer Systems},
pages = {59–75},
numpages = {17},
series = {EuroSys '25}
}

@INPROCEEDINGS{dvfs,
  author={Macken, P. and Degrauwe, M. and Van Paemel, M. and Oguey, H.},
  booktitle={IEEE International Conference on Solid-State Circuits}, 
  title={A voltage reduction technique for digital systems}, 
  year={1990},
  volume={},
  number={},
  pages={238-239},
}

@inproceedings{workload-dvfs,
author = {Lin, Chengdong and Wang, Kun and Li, Zhenjiang and Pu, Yu},
title = {A Workload-Aware DVFS Robust to Concurrent Tasks for Mobile Devices},
year = {2023},
booktitle = {Annual International Conference on Mobile Computing and Networking},
articleno = {19},
numpages = {16},
series = {MobiCom}
}

@article{learning-dvfs,
author = {Kim, Seyeon and Bin, Kyungmin and Ha, Sangtae and Lee, Kyunghan and Chong, Song},
title = {zTT: Learning-Based DVFS with Zero Thermal Throttling for Mobile Devices},
year = {2022},
issue_date = {December 2021},
volume = {25},
number = {4},
issn = {2375-0529},
journal = {GetMobile: Mobile Comp. and Comm.},
month = mar,
pages = {30–34},
}

@misc{cpu-dvfs,
  author       = {Rafael J. Wysocki},
  title        = {{CPU Performance Scaling}},
  year         = 2017,
  howpublished = {\url{https://docs.kernel.org/admin-guide/pm/cpufreq.html}},
  note         = {Accessed 25 Apr 2025},
}

@misc{lpddr5,
  author       = {JEDEC},
  title        = {LOW POWER DOUBLE DATA RATE (LPDDR) 5/5X},
  year         = 2023,
  howpublished = {\url{https://www.jedec.org/standards-documents/docs/jesd209-5c}},
  note         = {Accessed 25 Apr 2025},
}

@InProceedings{pmlr-v202-leviathan23a,
  title = 	 {Fast Inference from Transformers via Speculative Decoding},
  author =       {Leviathan, Yaniv and Kalman, Matan and Matias, Yossi},
  booktitle = 	 {Proceedings of the 40th ICML},
  pages = 	 {19274--19286},
  year = 	 {2023},
  volume = 	 {202},
  series = 	 {Proceedings of Machine Learning Research},
  month = 	 {23--29 Jul},
  publisher =    {PMLR},
}

@article{vaswani2023attentionneed,
  title={Attention is all you need},
  author={Vaswani, Ashish and Shazeer, Noam and Parmar, Niki and Uszkoreit, Jakob and Jones, Llion and Gomez, Aidan N and Kaiser, {\L}ukasz and Polosukhin, Illia},
  journal={Advances in neural information processing systems (NeurIPS)},
  volume={30},
  year={2017}
}

@article{gonzalez1997supply,
  title={{Supply and threshold voltage scaling for low power CMOS}},
  author={Gonzalez, Ricardo and Gordon, Benjamin M and Horowitz, Mark A},
  journal={IEEE Journal of Solid-State Circuits},
  volume={32},
  number={8},
  pages={1210--1216},
  year={1997},
}

@inproceedings{gpu-dvfs,
author = {Tang, Zhenheng and Wang, Yuxin and Wang, Qiang and Chu, Xiaowen},
title = {The Impact of GPU DVFS on the Energy and Performance of Deep Learning: an Empirical Study},
year = {2019},
booktitle = {Proceedings of the Tenth ACM International Conference on Future Energy Systems},
pages = {315–325},
numpages = {11},
}

@misc{oppo-npu,
    title = {{OPPO Find X8 Series to Debut MediaTek Dimensity 9400 SOC for Global Markets Combining Ultra Performance, Efficiency \& AI Experiences}},
    author = {OPPO},
    year = {2024},
    note         = {Accessed 25 Apr 2025}
}

@misc{open-llm-leaderboard,
  author       = {Edward Beeching and Clémentine Fourrier and Nathan Habib and Sheon Han and Nathan Lambert and Nazneen Rajani and Omar Sanseviero and Lewis Tunstall and Thomas Wolf},
  title        = {Open LLM Leaderboard},
  publisher    = {Hugging Face},
  howpublished = {\url{https://huggingface.co/spaces/HuggingFaceH4/open_llm_leaderboard}},
  note         = {Accessed 11 Apr 2025}
}

@software{mlc-llm,
    author = {{MLC team}},
    title = {{MLC-LLM}},
    year = {2023-2025}
}

@inproceedings{gshard,
    title={{GS}hard: Scaling Giant Models with Conditional Computation and Automatic Sharding},
    author={Dmitry Lepikhin and HyoukJoong Lee and Yuanzhong Xu and Dehao Chen and Orhan Firat and Yanping Huang and Maxim Krikun and Noam Shazeer and Zhifeng Chen},
    booktitle={International Conference on Learning Representations},
    year={2021}
}

@INPROCEEDINGS{edgetpu-predictor,
  author={Zou, Bohua and Sun, Binqi and Hu, Yigong and Kloda, Tomasz and Caccamo, Marco and Abdelzaher, Tarek},
  booktitle={IEEE Military Communications Conference (MILCOM)}, 
  title={A Performance Prediction-based DNN Partitioner for Edge TPU Pipelining}, 
  year={2024},
  pages={1-6},
}

@article{random-forest,
  title={Random forests},
  author={Breiman, Leo},
  journal={Machine learning},
  volume={45},
  pages={5--32},
  year={2001},
  publisher={Springer}
}

@article{mlp,
  title={Machine learning and deep learning},
  author={Janiesch, Christian and Zschech, Patrick and Heinrich, Kai},
  journal={Electronic markets},
  volume={31},
  number={3},
  pages={685--695},
  year={2021},
  publisher={Springer}
}

@article{svr,
  title={Support vector regression},
  author={Awad, Mariette and Khanna, Rahul and Awad, Mariette and Khanna, Rahul},
  journal={Efficient learning machines: Theories, concepts, and applications for engineers and system designers},
  pages={67--80},
  year={2015},
  publisher={Springer}
}

@article{poly,
    title = {Modelling using Polynomial Regression},
    journal = {Procedia Engineering},
    volume = {48},
    pages = {500-506},
    year = {2012},
    note = {Modelling of Mechanical and Mechatronics Systems},
    issn = {1877-7058},
    author = {Eva Ostertagová}
}

@misc{paul2025LLM-dvfs,
      title={Investigating Energy Efficiency and Performance Trade-offs in LLM Inference Across Tasks and DVFS Settings}, 
      author={Paul Joe Maliakel and Shashikant Ilager and Ivona Brandic},
      year={2025},
      eprint={2501.08219},
      archivePrefix={arXiv},
      primaryClass={cs.LG}, 
}

@inproceedings{ddl-dvfs,
author = {Yuan, Wanghong and Nahrstedt, Klara},
title = {Energy-efficient soft real-time CPU scheduling for mobile multimedia systems},
year = {2003},
isbn = {1581137575},
publisher = {Association for Computing Machinery},
address = {New York, NY, USA},
booktitle = {Proceedings of the Nineteenth ACM Symposium on Operating Systems Principles},
pages = {149–163},
numpages = {15},
location = {Bolton Landing, NY, USA},
series = {SOSP '03}
}

@ARTICLE{Mauricio2024LLM-ener,
  author={Argerich, Mauricio Fadel and Patiño-Martínez, Marta},
  journal={IEEE Access}, 
  title={Measuring and Improving the Energy Efficiency of Large Language Models Inference}, 
  year={2024},
  volume={12},
  number={},
  pages={80194-80207},
}

@misc{linux_pm_qos_doc,
  author       = {{The Linux Kernel Community}},
  title        = {Power Management Quality of Service (PM QoS) Interface},
  year         = {2024},
  note         = {Accessed 15 May 2025},
  howpublished = {\url{https://www.kernel.org/doc/html/latest/power/pm_qos_interface.html}}
}

@article{kendalltau,
    author = {KENDALL, M. G.},
    title = {A NEW MEASURE OF RANK CORRELATION},
    journal = {Biometrika},
    volume = {30},
    number = {1-2},
    pages = {81-93},
    year = {1938},
    month = {06},
    issn = {0006-3444},
}

@misc{team2024gemma,
      title={Gemma 2: Improving Open Language Models at a Practical Size}, 
      author={Gemma Team},
      year={2024},
      eprint={2408.00118},
      archivePrefix={arXiv},
      primaryClass={cs.CL}, 
}

@misc{aosp_overheat,
  author       = {{Google}},
  title        = {Thermal mitigation},
  year         = {2025},
  note         = {Accessed 15 May 2025},
  howpublished = {\url{https://source.android.com/docs/core/power/thermal-mitigation}}
}

@inproceedings{eprof,
author = {Pathak, Abhinav and Hu, Y. Charlie and Zhang, Ming and Bahl, Paramvir and Wang, Yi-Min},
title = {Fine-grained power modeling for smartphones using system call tracing},
year = {2011},
isbn = {9781450306348},
publisher = {Association for Computing Machinery},
address = {New York, NY, USA},
booktitle = {Proceedings of the Sixth Conference on Computer Systems},
pages = {153–168},
numpages = {16},
location = {Salzburg, Austria},
series = {EuroSys '11}
}

@article{readspeed-jornal,
title = {How many words do we read per minute? A review and meta-analysis of reading rate},
journal = {Journal of Memory and Language},
volume = {109},
pages = {104047},
year = {2019},
issn = {0749-596X},
author = {Marc Brysbaert},
}

@misc{intelpstate,
  author       = {{Rafael J. Wysocki}},
  title        = {intel pstate CPU Performance Scaling Driver},
  year         = {2017},
  note         = {Accessed 15 May 2025},
  howpublished = {\url{https://www.kernel.org/doc/html/latest/admin-guide/pm/intel_pstate.html}},
}

@misc{ollama,
  author       = {{Ollama}},
  title        = {Ollama: Chat \& build with open models},
  year         = {2025},
  note         = {Accessed 15 May 2025},
  howpublished = {\url{https://ollama.com/}},
}

@misc{hfl2023chinesellama,
  title        = {Chinese-LLaMA-2-1.3B: A Chinese-Enhanced LLaMA-2 Model},
  author       = {Harbin Institute of Technology and iFLYTEK Joint Laboratory (HFL)},
  year         = {2023},
  howpublished = {\url{https://huggingface.co/hfl/chinese-llama-2-1.3b}},
  note         = {Accessed 19 Aug 2025}
}

@inproceedings{chow2023cofris,
  title={CoFRIS: Coordinated frequency and resource scaling for GPU inference servers},
  author={Chow, Marcus and Wong, Daniel},
  booktitle={Proceedings of the 14th International Green and Sustainable Computing Conference},
  pages={45--51},
  year={2023}
}

@inproceedings{greathouse2018machine,
  title={Machine learning for performance and power modeling of heterogeneous systems},
  author={Greathouse, Joseph L and Loh, Gabriel H},
  booktitle={Proceedings of the International Conference on Computer-Aided Design},
  pages={1--6},
  year={2018}
}

@article{javaheripi2023phi,
  title={Phi-2: The surprising power of small language models},
  author={Javaheripi, Mojan and Bubeck, S{\'e}bastien and Abdin, Marah and Aneja, Jyoti and Bubeck, Sebastien and Mendes, Caio C{\'e}sar Teodoro and Chen, Weizhu and Del Giorno, Allie and Eldan, Ronen and Gopi, Sivakanth and others},
  journal={Microsoft Research Blog},
  volume={1},
  number={3},
  pages={3},
  year={2023}
}

@article{radford2019language,
  title={Language models are unsupervised multitask learners},
  author={Radford, Alec and Wu, Jeffrey and Child, Rewon and Luan, David and Amodei, Dario and Sutskever, Ilya and others},
  journal={OpenAI blog},
  volume={1},
  number={8},
  pages={9},
  year={2019}
}

@article{ainslie2023gqa,
  title={Gqa: Training generalized multi-query transformer models from multi-head checkpoints},
  author={Ainslie, Joshua and Lee-Thorp, James and De Jong, Michiel and Zemlyanskiy, Yury and Lebr{\'o}n, Federico and Sanghai, Sumit},
  journal={arXiv},
  year={2023}
}

@article{codrescu2014hexagon,
  title={Hexagon DSP: An architecture optimized for mobile multimedia and communications},
  author={Codrescu, Lucian and Anderson, Willie and Venkumanhanti, Suresh and Zeng, Mao and Plondke, Erich and Koob, Chris and Ingle, Ajay and Tabony, Charles and Maule, Rick},
  journal={IEEE Micro},
  volume={34},
  number={2},
  pages={34--43},
  year={2014},
}

@InProceedings{pmlr-v202-xiao23c,
  title = 	 {{S}mooth{Q}uant: Accurate and Efficient Post-Training Quantization for Large Language Models},
  author={Xiao, Guangxuan and Lin, Ji and Seznec, Mickael and Wu, Hao and Demouth, Julien and Han, Song},
  booktitle={International conference on machine learning (ICML)},
  pages={38087--38099},
  year={2023}
}

@article{kakolyris2024slo,
author = {Kakolyris, Andreas Kosmas and Masouros, Dimosthenis and Xydis, Sotirios and Soudris, Dimitrios},
title = {{SLO-Aware GPU DVFS for Energy-Efficient LLM Inference Serving}},
year = {2024},
volume = {23},
number = {2},
journal = {IEEE Computer Architecture Letters},
month = jul,
pages = {150–153},
numpages = {4}
}

@inproceedings{
zhang2026rethinking,
title={Rethinking {DVFS} for Mobile {LLM}s: Unified Energy-Aware Scheduling with {CORE}},
author={Zongpu Zhang and Pranab Dash and Qiang Xu and Y. Charlie Hu and Jian Li and Haibing Guan},
booktitle={MLSys},
year={2026},
url={https://openreview.net/forum?id=PSyHQ8kVUT}
}

@INPROCEEDINGS{edgeengine,
  author={Ahmadi, Amirhossein and Abdelhafez, Hazem A. and Pattabiraman, Karthik and Ripeanu, Matei},
  booktitle={2023 IEEE/ACM Symposium on Edge Computing (SEC)}, 
  title={EdgeEngine: A Thermal-Aware Optimization Framework for Edge Inference}, 
  year={2023},
  volume={},
  number={},
  pages={67-79}
  }
\end{document}